\pdfoutput=1
\documentclass[]{elsarticle}
\usepackage{lineno}
\usepackage{algpseudocode}
\usepackage{lipsum}
\usepackage{xcolor}
\usepackage{annotate-equations}
\usepackage{placeins}
\usepackage{amsmath,epsfig,subcaption,latexsym}
\usepackage{gensymb}
\usepackage{natbib}
\usepackage{mathtools,amssymb}
\biboptions{sort&compress}
\usepackage[justification=centering]{caption}
\usepackage{bm}
\usepackage{hyperref}
\usepackage{url}
\usepackage[export]{adjustbox}
\usepackage{textcomp}
\usepackage{array, tabularx,boldline}
\usepackage{cellspace}
\usepackage{xspace}

\newcommand{\flash}{FLASH\xspace}
\newcommand{\flashx}{Flash-X\xspace}
\newcommand{\amrex}{AMReX\xspace}
\newcommand{\paramesh}{Paramesh\xspace}
\newcommand{\hypre}{HYPRE\xspace}
\renewcommand{\Re}{\text{Re}}
\newcommand{\Fr}{\text{Fr}}
\renewcommand{\Pr}{\text{Pr}}
\newcommand{\St}{\text{St}}
\newcommand{\We}{\text{We}}
\newcommand{\Pe}{\text{Pe}}
\makeatletter
\newcommand{\vast}{\bBigg@{4}}
\newcommand{\Vast}{\bBigg@{5}}
\makeatother
\long\def\symbolfootnote[#1]#2{\begingroup%
\def\thefootnote{\fnsymbol{footnote}}\footnote[#1]{#2}\endgroup}

\newcount\ndots
\def\drwln#1#2{\raise 2.5pt\vbox{\hrule width #1pt height #2pt}}

\def\square   {${\vcenter{\hrule height .4pt
               \hbox{\vrule width .4pt height 3pt \kern 3pt
               \vrule width .4pt}
               \hrule height .4pt}}$\nobreak\ }

\def\filsqr   {${\vcenter{\hrule height 2pt
                        \hbox{\vrule width 2.2pt height 0.2pt \kern 0.1pt
                              \vrule width 2.2pt}
                              \hrule height 2.2pt}}$\nobreak\ }
\journal{Journal of Computational Physics}
\usepackage{multicol}
\begin{document}

\begin{frontmatter} 
\title{A Vortex Damping Outflow Forcing for Multiphase Flows with Sharp Interfacial Jumps}

\author[MCS]{Akash Dhruv \corref{cor2}}
\ead{} 
\address[MCS]{Mathematics and Computer Science Division,\\
Argonne National Laboratory, Lemont, IL, USA}
\cortext[cor2]{Corresponding author}

\begin{abstract}
{Outflow boundaries play an important role in multiphase fluid dynamics simulations that involve transition between liquid and vapor phases. These flows are dominated by low Weber numbers and a sharp jump in pressure, velocity, and temperature. Inadequate treatment of these jumps at the outlet generates undesirable fluid disturbances that propagate upstream and lead to instabilities within the computational domain. To mitigate these disturbances, we introduce a forcing term that can be applied to incompressible Navier-Stokes equations to enforce stability in the numerical solution. The forcing term acts as a damping mechanism to control vortices that are generated by droplet/bubbles in multiphase flows, and is designed to be a general formulation that can be coupled with a fixed pressure outflow boundary condition to simulate a variety of multiphase flow problems. We demonstrate its applicability to simulate pool and flow boiling problems, where bubble-induced vortices during evaporation and condensation present a challenge at the outflow. Validation and verification cases are chosen to quantify accuracy and stability of the proposed method in comparison to established benchmarks and reference solutions, along with detailed performance analysis for three-dimensional simulations on leadership supercomputing platforms. Computational experiments are performed using \flashx, which is a composable open-source software instrument designed for multiscale fluid dynamics simulations on heterogeneous architectures.}
\end{abstract}
%
\end{frontmatter}

%
%
\section{Introduction} \label{sc: introduction}
Outflow boundary conditions are used to enforce far-field behavior of flow variables in computational fluid dynamics (CFD) simulations, and are typically implemented to mitigate numerical instabilities caused by pressure fluctuations and turbulence at the exit of a computational domain. Their design and application are largely determined by nature of the problem, numerical formulation, limitations of the computational box size, and their effect on the overall accuracy and reliability of the solution. Present literature covers a variety of boundary conditions for the simulation of single and multiphase incompressible flows \cite{Dong2014a,Dong2014,BOZONNET2021110528,POUX20114011,LIU20097250,BLAYO2005231}, each with their advantages and drawbacks when coupling pressure and velocity to enforce momentum and continuity equations.

Within the context of a fractional step, staggered-grid, predictor-corrector (projection) method for the incompressible Navier-Stokes equations, the simplest implementation involves setting a fixed-pressure outflow, and forcing velocity-gradients normal to the boundary to be zero. This approach implicitly satisfies mass conservation through coupling between predictor, pressure Poisson, and corrector equations that are solved in a sequence to enforce momentum and continuity constraints; and is suitable for flows without an explicitly defined inlet or with uniform pressure distribution at far-field boundaries, as in large-scale atmospheric simulations.

For turbulent flows with an explicitly defined inlet, such as flow over an aircraft geometry, a different approach is used. This involves setting a zero pressure-gradient at the outflow, accompanied by solving a convection equation for velocity at the boundary during the corrector step. The convective velocity of the equation is set equal to the maximum or mean velocity of the flow at the outflow boundary, which forces the vortices to exit the domain without generating a backflow. Due to the nature of the implementation, that is, replacing the solution of incompressible Navier-Stokes equations with a convective transport equation, explicit mass conservation is enforced by scaling the velocities at the outflow based on the mean velocity at the inlet. This popular treatment was implemented by Vanella et al. for single-phase fluid-structure interaction simulations within \flash \cite{Vanella2009}, a multiphysics simulation software framework.

Outflow treatments in general influence the flow behavior upstream and introduce bias in numerical solution. However, they serve has as a more favorable alternative to using larger simulations domains which increase the computational cost, and help mitigate numerical instabilities that arise from non-physical backflow propagations. As a result, this topic has garnered significant attention from researchers in various domains to design formulations that reduce outflow-induced errors in the numerical solution. The most sophisticated approach involves imposing a traction boundary condition that is derived from the weak form of Navier-Stokes equations \cite{taylor_rance_medwell, BOZONNET2021110528}, and applied to modify the formulation of the Pressure Poisson Equation (PPE) for the fractional-step projection method for incompressible Navier-Stokes equations \cite{LIU20097250,POUX20114011,BOZONNET2021110528}.

Extension of outflow boundary condition to multiphase flows requires special considerations to satisfy mass conservation for individual phases, and mitigate pressure and velocity fluctuations due to interfacial jumps. Traction boundary conditions are excellent candidates for designing a general solution to model outflow in multiphase flow simulations, however, the complexity lies in designing appropriate traction force that can balance effect of various multiphase dynamics. Many have investigated this problem in detail. Dong et al. \cite{Dong2014a,Dong2014} have made contributions in this area for flows involving multiphase jets with explicitly defined inlet and outlet. Their formulation is based on a phase field approach to interface tracking with a mollified forcing for multiphase boundary conditions, which is applicable to the simulation of flows where inertial force dominates surface tension, resulting in high Weber numbers ($\text{We} > 10^3$). Recently, Bozonnet et al. \cite{BOZONNET2021110528} made advances in this area by designing a formulation that allows for stability in multiphase flow simulations where backflow is a natural solution.

For flows involving phase transition, where inertial and surface tension forces are in a balance ($\text{We} = 1$) new challenges related to mass conservation and sharp treatment of interfacial jumps have emerged. These challenges particularly arise due to large surface tension forces at the triple contact point of liquid, vapor and outflow boundary, which lead to instability in the pressure solver, eventually affecting the divergence of velocity and overall solution of momentum equations. In addition, local effect of evaporation poses difficulties with explicit mass balance within the domain, presenting a new research avenue to develop outflow boundary condition for these types of problems.

{Numerical simulations of multiphase flows with sharp interfacial jumps have been conducted by many researchers in the last decade \cite{Gibou2007,Son2008,Mukherjee2004,SATO2017505,DHRUV2019,DHRUV2021,SHIN2002427,TRYGGVASON2006660}, with contributions ranging from fundamental mathematical formulations to production simulations that mimic experimental setups and provide new insights into the physics of bubble/droplet interactions. Considerable effort has also been made to model phase-transition. These efforts range from developing numerical methods \cite{SonDhir,MALAN2021109920,NINGEGOWDA2020120382} and benchmarks \cite{TANGUY20141} to performing complex simulations to gain physical insights into the process of boiling across a range of parametric space \cite{URBANO2019118521,SATO2018876}}. In our previous work \cite{DHRUV2019}, we extended \flash to focus on the problem of pool boiling and designed an adaptive mesh refinement (AMR)-based solver for high-fidelity simulations, and performed an extensive numerical study to validate and quantify effects of gravity on boiling heat transfer mechanisms \cite{DHRUV2021}. The outflow boundary condition consisted of a fixed pressure treatment with normal velocity gradients set to zero at the boundary. To avoid numerical instabilities at the outflow due to multiphase interactions described above, we introduced artificial condensation for bubbles based on the treatment proposed by Sato et al. \cite{Sato2013} which leads to complete condensation from vapor to liquid within a buffer region near the outflow. This treatment mimicked a type of sponge/nudging layer that has been widely used to damp flows widely in oceanic simulations \cite{BLAYO2005231}. As we tried to exercise our solver for a range of different cases for pool and flow boiling simulations, spanning a parametric space of different degrees of subcooling, wall superheat, and vapor-liquid properties, numerical challenges started to appear. The buffer region was characterized by its length and the strength of artificial subcooling. To completely condense vapor bubbles into liquid, the length and strength of the buffer region had to be adjusted on the basis of the degree of subcooling, gravity, and the rest of the parametric space. This became heuristic and often tedious process when deploying production simulations. Additionally, artificial condensation introduced non-physical vortices that necessitated development of a more formal treatment.

Extending traction boundary conditions for phase transition problems in our AMR based solver poses several challenges. Existing multiphase formulations are implemented using a variable coefficient PPE \cite{BOZONNET2021110528}. Within our framework, we use a multigrid Poisson solver that would require building the coefficient matrix associated with this PPE every time step as the multiphase system evolves. {For computational efficiency, we solve a constant coefficient PPE for the variable-density system based on the methodology described in \cite{DODD2014416}}. Extending multiphase traction boundary condition is not trivial for this approach. Additionally, using homogeneous boundary conditions for the pressure solver and formalizing the sponge/nudging layer based approach can serve as a foundation for developing more robust traction boundary conditions in the future. Therefore, in this work, we focus on treating outflow in multiphase flows using a forcing term for velocity and pressure gradients. The formulation is implemented for a sharp interface ghost fluid method (GFM) but can be easily extended to other interface capturing methods with minor modifications. The forcing term is used with a fixed pressure outflow condition to mitigate numerical instabilities that arise in problems like pool and flow boiling. The numerical framework is implemented within \flashx \cite{DUBEY2022}, a multiphysics simulation software instrument derived from \flash, and is actively being developed for performance portability and heterogeneous computing.

The paper is structured as follows, Section \ref{sc:governing-equations} provides an overview of the mathematical formulation and governing equations, followed by Section \ref{sc:numerical-formulation} which provides details of numerical implementation and performance metrics. Section \ref{sc:outflow-forcing} discusses the outflow forcing, and finally results and conclusions are provided in Sections \ref{sc:results} and \ref{sc:conclusion}, respectively.
%
%
\section {Governing Equations} \label{sc:governing-equations}
Figure \ref{fig:schematic} provides a schematic of pool and flow boiling simulations that serve as fundamental tools to understand design of thermal cooling systems for industrial, automotive, and electronic applications. Boiling occurs when a liquid coolant undergoes evaporation on the surface of a solid heater, resulting in formation of gas (vapor) bubbles which induce turbulence and improve heat transfer efficiency. The dynamics of the bubbles is governed by the balance between gravity ($g$), surface tension ($\sigma$) and evaporative heat flux. In the simulation, the liquid-gas interface, $\Gamma$, is tracked using a level-set, $\phi$, a signed distance function that is positive inside the gas and negative inside the liquid. $\phi = 0$, represents the implicit location of $\Gamma$.

During flow boiling, the liquid coolant flows over the heater surface with a bulk temperature, $T_b$ ($T_{bulk}$) and a velocity, $U_b$. The heater acts as a no-slip wall with constant temperature, $T_w$ ($T_{wall}$), and promotes the evaporation and formation of bubbles. The corresponding boundary conditions for velocity ($u$), pressure ($P$) and temperature ($T$) are shown in Figure \ref{fig:schematic}a using tensor notation to represent derivatives and direction.

The departure of bubbles from the heater surface and their subsequent movement through the liquid produce vortices that induce turbulence and circulatory flow within the computational domain. Vortices are the result of interfacial dynamics related to evaporation, condensation, and surface tension, which also appear during pool boiling. Pool boiling is similar to flow boiling without any liquid inflow. See the boundary conditions in Figure \ref{fig:schematic} b for more details.

{ The dynamics of boiling are modeled using a coupled system of momentum and energy equations. Assuming an incompressible flow, the equations can be described for each phase using tensor notations for three dimensional space. The equations for liquid phase are written as,
\begin{subequations} \label{eq:transport-liq}
\begin{equation} \label{eq:momt-liq}
\frac{\partial u_i}{\partial t} + u_j \partial_j u_i = - \frac{1}{\rho_L'} \partial_i P + \frac{1}{\rho_L'\:\Re} \partial_j \Big(\mu_L'\partial_j u_i + \mu_L'\partial_i u_j\Big) + \frac{g_i}{\Fr^2} 
\end{equation}
\begin{equation} \label{eq:temp-liq}
\frac{\partial T}{\partial t} + u_j \partial_j T = \frac{1}{\rho_L' C_{p_L}' \Pe} \partial_j \Big(k_L' \partial_j T\Big)
\end{equation}
\end{subequations}
And for the gas phase as,
\begin{subequations} \label{eq:transport-gas}
\begin{equation} \label{eq:momt-gas}
\frac{\partial u_i}{\partial t} + u_j \partial_j u_i = - \frac{1}{\rho_G'} \partial_i P + \frac{1}{\rho_G'\:\Re} \partial_j \Big(\mu_G'\partial_j u_i + \mu_G'\partial_i u_j\Big) + \frac{g_i}{\Fr^2} 
\end{equation}
\begin{equation} \label{eq:temp-gas}
\frac{\partial T}{\partial t} + u_j \partial_j T = \frac{1}{\rho_G' C_{p_G}' \Pe} \partial_j \Big(k_G' \partial_j T\Big)
\end{equation}
\end{subequations}

\begin{figure}[h]
    \centering
    \includegraphics[width=0.9\textwidth]{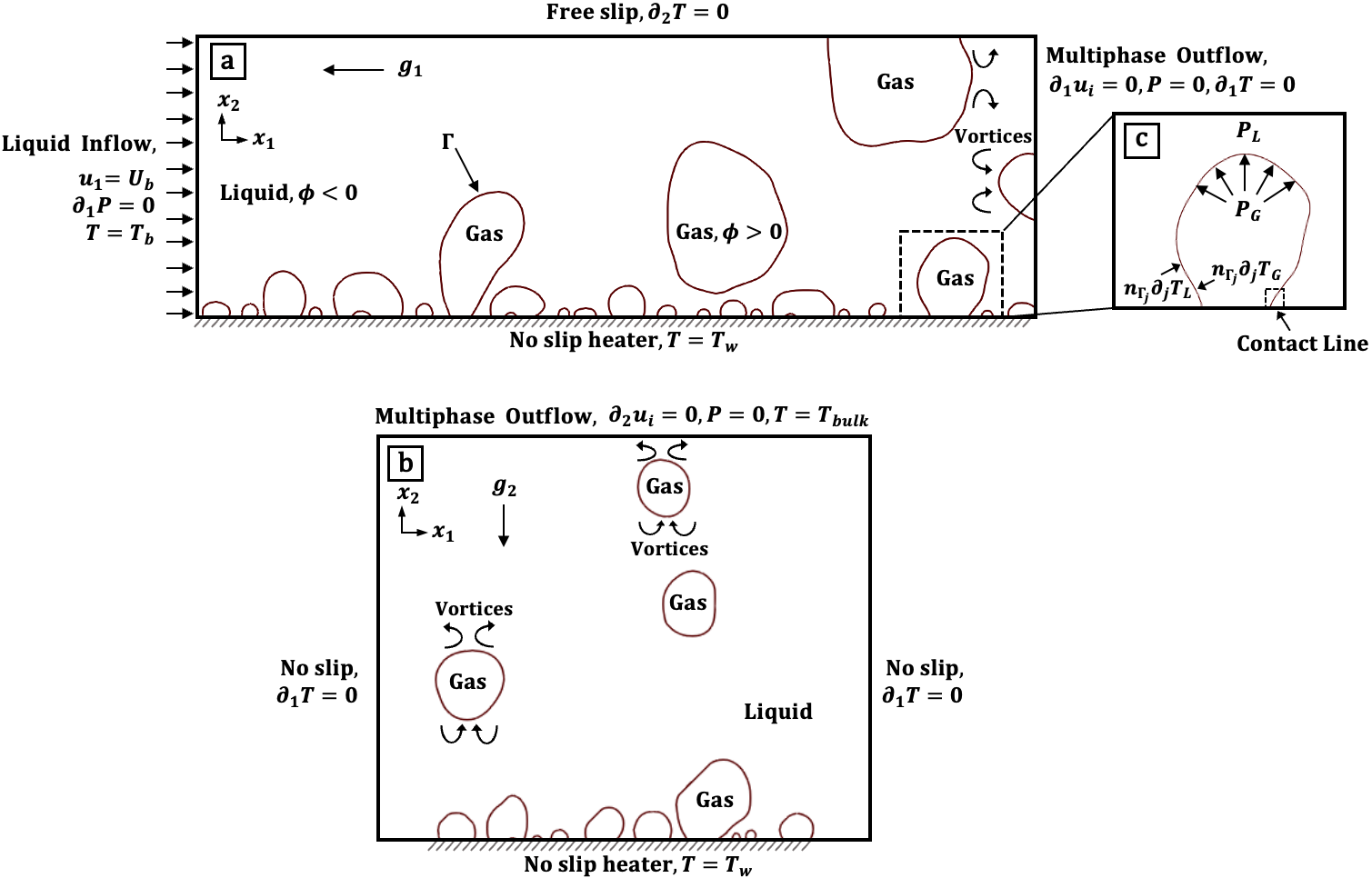}
    \caption{Schematic of multiphase heat transfer problems: (a) Flow boiling (b) Pool boiling, showing vortices generated by gas bubbles along with boundary conditions for velocity ($u_i$), pressure ($P$), and temperature ($T$). The level-set function ($\phi$) implicitly tracks the liquid-gas interface ($\Gamma$). The inset (c) provides details of the contact line, the pressure jump, and the evaporative mass flux.}
\label{fig:schematic}.
\end{figure}

$u$ is the velocity, ${P}$ is the pressure, and ${T}$ is the temperature everywhere in the domain. The equations are non-dimensionalized with reference quantities from the liquid phase resulting in Reynolds number (${\Re}$) $ = \rho_L u_0 l_0/\mu_L$, Prandtl number (${\Pr}$) $ = \mu_L C_{p_L}/k_L$, Froude number (${\Fr}$) $ = u_0/\sqrt{gl_0}$, and Peclet number (${\Pe}$) $ = \Re\:\Pr$. $u_0$ and $l_0$ are reference velocity and length scales, and $g$ is acceleration due to gravity. The symbols $\rho_L$, $\mu_L$, $C_{p_L}$ and $k_L$ represent density, viscosity, specific heat and thermal conductivity of the liquid. For boiling problems, the reference length scale is set to the capillary length $l_0 = \sqrt{\sigma/(\rho_L-\rho_G) \: g}$ (where $\sigma$ is the surface tension), and the velocity scale is the terminal velocity $u_0=\sqrt{g\:l_0}$. The reference temperature scale is given by $(T-T_{bulk})/\Delta T$, where $\Delta T = T_{wall} - T_{bulk}$. Additionally, the non-dimensional fluid properties in each phase are given by,

\begin{equation} \label{eq:ratios}
y_L' =  1,  \quad y_G' =  \frac{y_G}{y_L}
\end{equation}

where $y$ corresponds to $\rho$ (density), $\mu$ (viscosity), $C_p$ (specific heat), and $k$ (thermal conductivity).}

{The momentum equations strictly follow incompressible formulation reducing the viscous term in Equation \ref{eq:transport-liq}a and \ref{eq:transport-liq}b to, 
\begin{equation}
\partial_j \Big(\mu'\partial_j u_i + \mu'\partial_i u_j \Big)  = \partial_j \Big(\mu'\partial_j u_i \Big) + \mu'\partial_i \Big(\partial_j u_j\Big) = \partial_j \Big(\mu'\partial_j u_i \Big)
\end{equation}
where $\mu'$ is $\mu_L'$ or $\mu_G'$ depending on the phase.

This reduction is based on the constraint that divergence of velocity is zero in both phases. The effect of evaporation, which leads to non-zero divergence near the interface, is handled separately using interfacial jump conditions discussed further in the text. Unless specified otherwise, physical quantities are assumed to be non-dimensional in the following discussions. Additionally, Einstein's convention for summation is observed for repeating tensor indices. In special cases where this convention is violated—specifically, when repeating indices appear more than twice in a term—an explicit summation operator is provided.}

{The continuity equation is given by,

\begin{equation} \label{eq:continuity}
\partial_j  u_j = - \dot m \: n_{\Gamma_j} \partial_j  \frac{1}{\rho'} \:\Biggr\vert_{\:\Gamma}
\end{equation}

where, {$n_{\Gamma_i} = \partial_i \phi/\sqrt{\partial_j \phi \partial_j \phi}$}}, is a unit vector normal to the liquid-gas interface, and, $\dot m$, is the evaporative mass flux. It is important to emphasize that the fluid flow is fully incompressible within each phase, and the right-hand side of Equation \ref{eq:continuity} accounts for the mass balance due to the change in phase from liquid to gas and vice versa. Furthermore, $\dot m \: n_{\Gamma_j} \partial_j  \frac{1}{\rho'}$, is zero everywhere {except} at the interface. 

Figure \ref{fig:schematic}c shows a schematic of the heat flux from liquid to gas, $n_{\Gamma_j} \partial_j T_L$ and the heat flux from gas to liquid, $n_{\Gamma_j} \partial_j T_G$, normal to the liquid-gas interface, $\Gamma$. The difference between these heat fluxes determines the mass transfer from one phase to the other. {The value of ${\dot m}$ at the interface is calculated as,

\begin{equation} \label{eq:evaporation}
\dot m = \frac{\St}{\Re\:\Pr} \Bigg( {n_{\Gamma_j} k_L' \: \partial_j T_L \:\bigr\vert}_{\:\Gamma} - n_{\Gamma_j} k_G' \: {\partial_j T_G  \:\bigr\vert}_{\:\Gamma} \Bigg)
\end{equation}

where ${\St}$ is the Stefan number given by,  $\St = C_{p_L} \Delta T/Q_l$, with ${Q_l}$ as latent heat of evaporation.} The heat flux from the liquid region, $\partial_j T_L$, contributes to evaporation, and the heat flux from the gas region, $\partial_j T_G$, accounts for condensation. 

{The level-set function representing the interface is computed using the convection equation,

\begin{equation} \label{eq:levelset-convection}
\frac{\partial \phi}{\partial t} + u_{\Gamma_j} \partial_j \phi = 0,
\end{equation}}

where, $u_{\Gamma_i} = u_i + (\dot m/\rho')n_{\Gamma_i}$, is the interface velocity. The convection of level-set is accompanied by a selective reinitialization technique to mitigate diffusive errors that are generated from numerical descretization of the convection term. The detailed approach for this is discussed in \cite{DHRUV2019,akash_phd_2021}

{Reorganizing equation for, $u_{\Gamma_i} = u_i + (\dot m/\rho')n_{\Gamma_i}$, results in the following expression for jump in velocity normal to the liquid-gas interface:

\begin{equation} \label{eq:vel-jump}
[u]_{\Gamma_i} = u_{G_i} - u_{L_i} = \dot m \: n_{\Gamma_i} \Bigg[ \frac{1}{\rho_G'} -  \frac{1}{\rho_L'} \Bigg]
\end{equation}}

The mass flux, $\dot m$, surface tension, $\sigma$, and viscous stresses contribute towards a similar jump in pressure: where, ${\kappa}$ is the interface curvature. The non-dimensional form for pressure jump using, ${\We} = \rho_l u_0^2 l_0/\sigma$, is consequently given by:

{\begin{equation} \label{eq:pres-jump}
[P]_{\Gamma} = P_g-P_l = \frac{\kappa}{\We} - \Bigg[ \frac{1}{\rho_G'} -  \frac{1}{\rho_L'} \Bigg] \dot m^2
\end{equation}}

{ The effect of viscous jump is assumed to negligible in this formulation due to the smeared treatment of viscosity near the interface described in the following section. This assumption fits with the formulation in \cite{Kang2000}.}

Finally, the boundary condition for temperature at the liquid-gas interface is given by,

\begin{equation} \label{eq:temp-jump}
T_{\Gamma}  = T_{sat}
\end{equation}

where, $T_{sat}$ corresponds to saturation temperature. 

{ The jump conditions in Eqs \ref{eq:vel-jump}, \ref{eq:pres-jump}, \ref{eq:temp-jump} are modeled using a GFM \cite{Gibou2007,DHRUV2019,akash_phd_2021}.}

\section {Numerical Formulation} \label{sc:numerical-formulation}
\begin{figure}[h]
    \centering
    \includegraphics[width=0.4\textwidth]{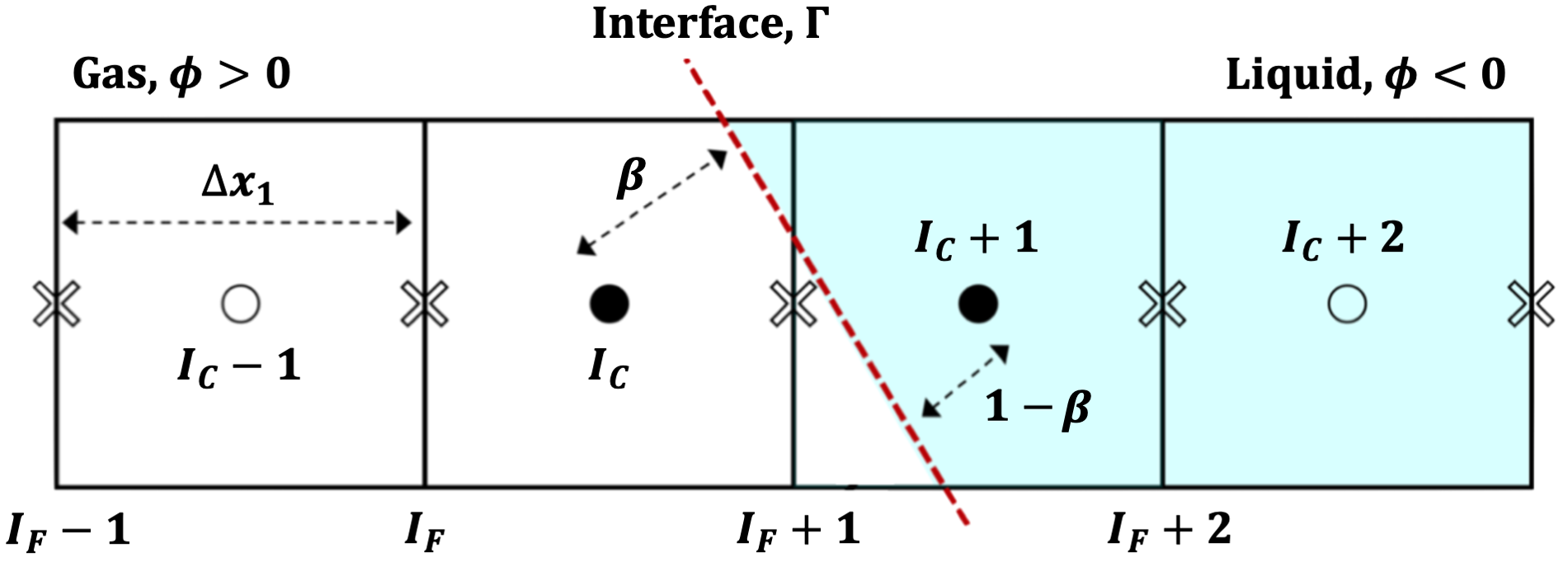}
    \caption{Staggered computational grid highlighting faces ($\times$) and cell-centers ($\bullet$), $I_C$ denotes the cell-centers and $I_F$ if face-center. The liquid-gas interface, $\Gamma$, separates the two phases.}
    \label{fig:stencil_staggered}
\end{figure}

{For the purposes of numerical formulation Eqs \ref{eq:transport-liq}, \ref{eq:transport-gas}, and \ref{eq:vel-jump}-\ref{eq:temp-jump} are merged into a single-fluid approach, which has variable properties in space (either liquid or gas) depending on the location of the interface. These equations are discretized on a block-structured Cartesian grid in the full domain containing both phases. The merged equations are written as follows,
\begin{subequations} \label{eq:transport}
\begin{equation} \label{eq:momt}
\frac{\partial u_i}{\partial t} + u_j \partial_j u_i = \frac{1}{\rho'} \Big(-\partial_i P + \text{H}_\Gamma\text{S}^{\Gamma}_{\partial_i P} + \text{S}^{\text{O}}_{\partial_i P}) + \frac{1}{\rho'\:\Re} \partial_j \Big(\mu'\partial_j u_i + \mu'\partial_i u_j\Big) + \frac{g_i}{\Fr^2} + \text{H}_\Gamma\text{S}^{\Gamma}_{u_i}  + \text{S}^{\text{O}}_{u_i} 
\end{equation}
\begin{equation} \label{eq:temp}
\frac{\partial T}{\partial t} + u_j \partial_j T = \frac{1}{{\Pe}}\partial_j \Big(\alpha' \partial_j T\Big) + \text{H}_\Gamma\text{S}^{\Gamma}_{T}
\end{equation}
\end{subequations}
where thermal diffusivity $\alpha' = k'/(\rho'C_p')$ is used to consolidate the the diffusion term for the energy equation. The effects of interfacial jumps at $\Gamma$ ($\phi=0$) in Eqs \ref{eq:vel-jump}, \ref{eq:pres-jump}, and \ref{eq:temp-jump} are given by the source terms $\text{S}^{\Gamma}_{u_i}$, $\text{S}^{\Gamma}_{\partial_i P}$, and $\text{S}^{\Gamma}_{T}$ respectively. These terms are only relevant when $\text{H}_\Gamma=1$. Based on $\phi$, the value of $\text{H}_\Gamma$ is 1 only in the cells near $\phi=0$ and is zero otherwise. The source terms, $\text{S}^{\text{O}}_{u_i}$ and $\text{S}^{\text{O}}_{\partial_i P}$, correspond to the treatment of the outflow which is the main focus of this work.

A finite difference spatial and temporal discretization is implemented for Eqs \ref{eq:transport} - \ref{eq:temp-jump}, representing a full system for multiphase fluid dynamics equations solution. The vectors along with $\rho'$ are defined at face-centers, and the scalars along with remaining properties are defined at cell-centers. The non-dimensional fluid properties at cell-centers are given as,

\begin{equation} \label{eq:properties}
\mu' =  \mu_L'\:(1-\text{H}_\epsilon) + \mu_G'\:\text{H}_\epsilon, \quad \alpha' =  \alpha_L'\:(1-\text{H}_\epsilon) + \alpha_G'\:\text{H}_\epsilon
\end{equation}

where $\text{H}_\epsilon$ is a smeared Heaviside function,
\begin{equation}
\text{H}_\epsilon = \begin{cases}
     0,& \text{if} \: \phi < -\epsilon \\
     \frac{1}{2}[1+\frac{\phi}{\epsilon}+\frac{1}{\pi}\text{sin}(\frac{\pi\phi}{\epsilon})],& \text{if} \: |\phi| \leq \epsilon \\
     1,& \text{if} \: \phi > \epsilon\\
\end{cases}
\end{equation}

and $\epsilon$ is a tolerance that is chosen as $1.5$ times the grid spacing. 

The face-centered density is defined as,
\begin{equation}
\rho' =  \rho_L'\:(1-\text{H}) + \rho_G'\:\text{H},  \quad 
\end{equation}

where $\text{H}$ is a sharp Heaviside function,
\begin{equation} \label{eq:heaviside}
\text{H} = \begin{cases}
     0,& \text{if} \: \phi < 0 \\
     1,& \text{if} \: \phi \geq 0\\
\end{cases}
\end{equation}

Additionally, a special smearing strategy is applied for face-centered density. Consider Figure \ref{fig:stencil_staggered}, which describes the computational stencil in direction $x_1$ with cell-centers ($I_C$) and faces ($I_F$). Near the liquid gas interface, $\Gamma$, the density is smeared using,

\begin{equation} \label{eq:smeared-density}
\rho' \Bigr|_{I_F+1} =  \rho_L'\:(1-\beta) + \rho_G'\:\beta
\end{equation}

where, $\beta = {|\phi^{n+1}_{I_c}|}\Big/\Big({|\phi^{n+1}_{I_c}|+|\phi^{n+1}_{I_c+1}|}\Big)$.}

{The advection terms are calculated using a fifth-order Hamilton-Jacobi weighted essentially non-oscillatory (WENO) method described in \cite{doi:10.1137/S106482759732455X, kpd_thesis}, and the diffusion terms are calculated using a second-order central difference discretization}. Temporal integration of the convective and diffusion terms is implemented using a second-order Adams-Bashforth scheme.

The momentum Eq \ref{eq:momt} follows a fractional-step predictor-corrector approach to enforce continuity in Eq \ref{eq:continuity}. The multistep method is given as follows,

\begin{equation} \label{eq:predictor}
\frac{u^*_i - u^n_i}{\Delta t} = \frac{3}{2}\Theta^n_i - \frac{1}{2}\Theta^{n-1}_i + \frac{g_i}{\Fr^2} + \text{S}^{\Gamma}_{u_i} + \text{S}^{\text{O}}_{u_i} 
\end{equation}

Eq \ref{eq:predictor} is the predictor step that does not include contribution from pressure or surface tension forces. $\Theta_i$, represents the contribution of advection and diffusion terms, and $n$ denotes the solution in time as the simulation progresses with a time step, $\Delta t$. Assuming, $\zeta^{n+1}_i = \partial_i P^{n+1} - \text{S}^{\Gamma}_{\partial_i P} - \text{S}^{\text{O}}_{\partial_i P}$, the corrector equation becomes,

\begin{subequations} \label{eq:corrector}
\begin{equation} \label{eq:corrector-a}
\frac{u^{n+1}_i - u^*_i}{\Delta t} = - \frac{1}{\rho_G'} \zeta^{n+1}_i - \frac{1}{\rho'} (2\zeta^n_i - \zeta^{n-1}_i) + \frac{1}{\rho_G'} (2\zeta^n_i - \zeta^{n-1}_i)
\end{equation}
{\begin{equation} \label{eq:corrector-b}
\frac{u^{n+1}_i - u^*_i}{\Delta t} = - \frac{1}{\rho_G'} \partial_i P^{n+1} + \frac{1}{\rho_G'} (\text{S}^{\Gamma}_{\partial_i P} + \text{S}^{\text{O}}_{\partial_i P}) + \Bigg(\frac{1}{\rho_G'} - \frac{1}{\rho'} \Bigg) (2\zeta^n_i - \zeta^{n-1}_i)
\end{equation}}
\end{subequations}

{The splitting of the pressure term is implemented to convert the variable coefficient Poisson equation into a configuration with constant coefficients. This splitting is based on the formulation described in \cite{DODD2014416}, which separates the pressure term into a constant term (the unknown) and a variable term that utilizes linear extrapolation from time steps $n$ and $n-1$. Consequently, the divergence of the corrector step in Equation \ref{eq:corrector-b} leads to the following pressure equation,

\begin{equation} \label{eq:poisson}
\partial_j \partial_j P^{n+1} = \frac{\rho_G'}{\Delta t} \Bigg(\partial_j u^*_j - \partial_j u^{n+1}_j\Bigg) + \sum_{j=1}^{3} \partial_j \Bigg(1 - \frac{\rho_G'}{\rho'} \Bigg) (2\zeta^n_j - \zeta^{n-1}_j)  \\
+ \partial_j \text{S}^{\Gamma}_{\partial_j P} + \partial_j \text{S}^{\text{O}}_{\partial_j P}
\end{equation}

$\partial_j u^{n+1}_j$ is calculated using Equation \ref{eq:continuity}.} In reference to Figure \ref{fig:stencil_staggered}, the source terms $\text{S}^{\Gamma}_{u_1}$ and $\text{S}^{\Gamma}_{\partial_1 P}$ are given as,

{
\begin{subequations} \label{eq:source-vel-pres}
\begin{equation}
\text{S}^{\Gamma}_{u_1} \Bigr|_{I_F+1} = \frac{1}{\rho'\:\Re} \partial_1 \Big[\mu' \partial_1 (\hat{u}_1 - u_1) \Big] \: \Biggr|_{I_F+1}
\end{equation}
\begin{equation}
\text{S}^{\Gamma}_{\partial_1 P} \Bigr|_{I_F+1} = \frac{\kappa^{n+1}_{I_c} (1-\theta) + \kappa^{n+1}_{I_c+1} \theta}{\text{We} \: \Delta x_1} - \Bigg(\frac{1}{\rho_G'} - \frac{1}{\rho_L'} \Bigg)\frac{\dot{m^2}^{n+1}_{I_c} (1-\theta) + \dot{m^2}^{n+1}_{I_c+1} \theta}{\Delta x_1}
\end{equation}
\end{subequations}
where curvtature, $\kappa = \partial_j n_{\Gamma_j}$, and $\hat{u}_1$ is the ghost fluid velocity that is applied to the viscous term in Equation \ref{eq:transport}a. Using the definition of interface velocity in Equation \ref{eq:vel-jump}, $\hat{u}_1 = u_1 + (\dot m/\rho')n_{\Gamma_1}$, modifies Equation \ref{eq:source-vel-pres}a to,

\begin{equation} \label{eq:source-vel}
\text{S}^{\Gamma}_{u_1} \Bigr|_{I_F+1} = \frac{1}{\rho'\:\Re} \partial_1 \Bigg(\frac{\mu'\dot{m}}{\rho'}\partial_1 n_{\Gamma_1} \Bigg) \: \Biggr|_{I_F+1}
\end{equation}

The corresponding discretization of the energy and level-set equations is discussed at length in \cite{DHRUV2019,akash_phd_2021}, along with the verification of the overall second-order accuracy of the multiphase solver. Calculation of the forcing term $\text{S}^{\Gamma}_{T}$ is similar to formulation of  $\text{S}^{\Gamma}_{u_1}$ for the stencil in Figure \ref{fig:stencil_staggered},

\begin{equation} \label{eq:source-temp}
\text{S}^{\Gamma}_{T} \Bigr|_{I_{C+1}} = \frac{1}{\Pe}\partial_1 \Big[{\alpha'}\partial_1 (\hat{T} - T) \Big] \: \Biggr|_{I_{C+1}}
\end{equation}

In this case, $\hat{T}$ is the ghost fluid temperature which is applied to the diffusion term in Equation \ref{eq:transport}b and calculated as,

\begin{equation} \label{eq:temp-gfm}
\hat{T}_{I_{C+1}} = T_{I_{C}} + \Delta x_1 \frac{T_\Gamma - T_{I_{C}}}{\theta \Delta x_1}
\end{equation}

where $T_\Gamma = T_{sat}$ from Equation \ref{eq:temp-jump}. Ghost fluid boundary conditions are only applied to diffusion terms, while the WENO advection scheme effectively preserves the sharp jump in properties and does not require special treatment.}

\begin{figure}[h]
    \centering
    \includegraphics[width=0.8\textwidth]{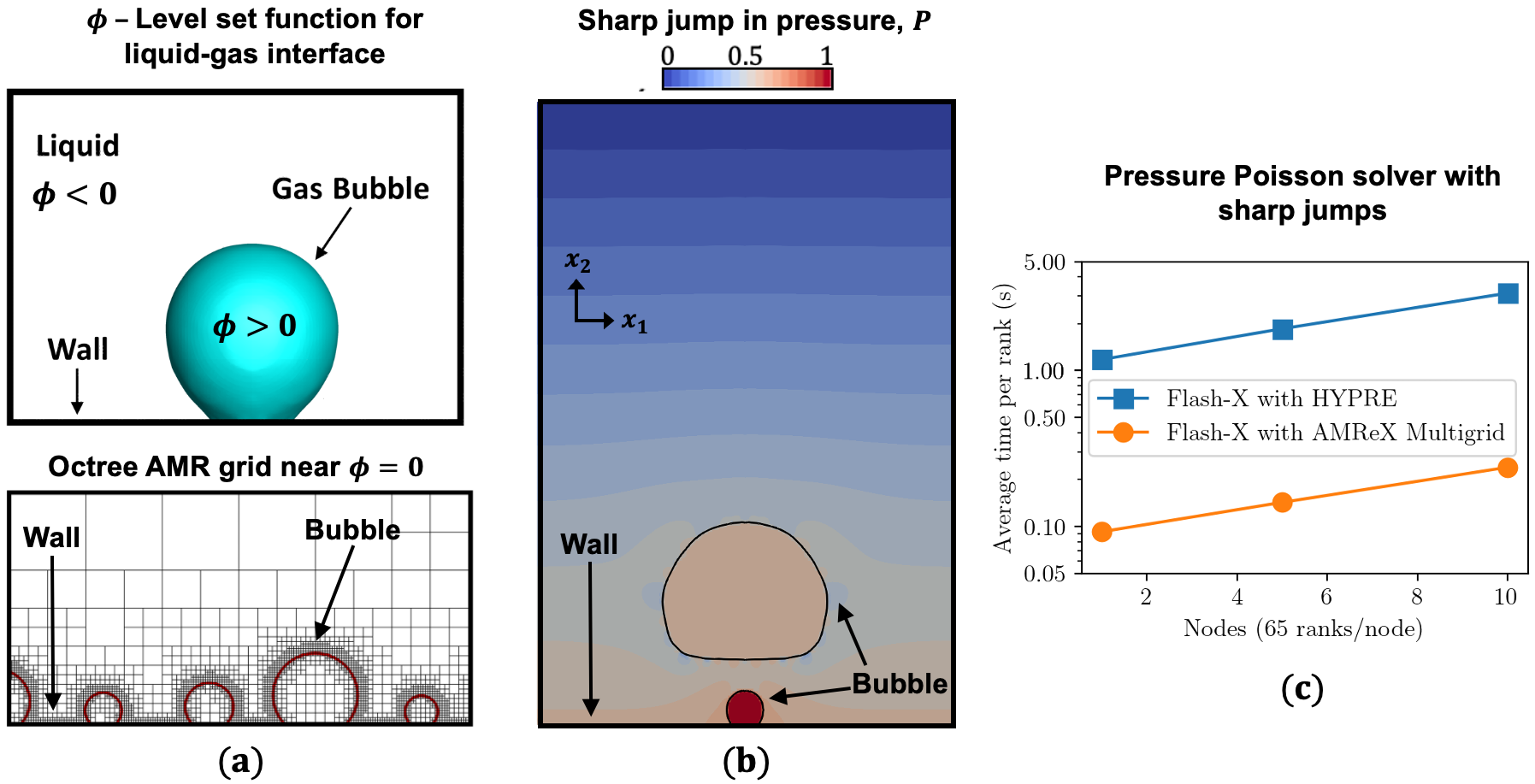}
    \caption{(a) Schematic of a multi-level block structured AMR grid around the liquid-gas interface, $\phi=0$, (b) Distribution of scaled pressure, $\frac{P-P_{low}}{P_{high}-P_{low}}$, highlighting the jump in value between phases, and (c) Weak scaling of pressure Poisson solver on Summit \cite{summit}}
    \label{fig:performance}
\end{figure}

Block-structured, octree-based adaptive mesh refinement (AMR) is used to selectively refine the grid in the regions of interest. The default AMR backend within \flash was implemented using \paramesh along with a customized multigrid Poisson solver that leverages \hypre to calculate the solution at the coarsest refinement level. \flashx, on the other hand has \amrex as an additional option for the management of the AMR network. \amrex \cite{AMReX_JOSS}, which is natively a patch-based  library, can also mimic the octree mode when needed. We use the octree mode of \amrex for a direct comparison of the performance of the two AMR backends and their multigrid solvers in multiphase fluid dynamics applications within \flashx. Figure \ref{fig:performance}a provides a schematic of AMR around the liquid-gas interface located at $\phi=0$. The refinement criteria is set to satisfy the condition $\lvert\phi\rvert \leq \epsilon $, where $\epsilon$ is the tolerance that sets a range for $\phi$ to provide thresholds for refinement and derefinement. 

Figure \ref{fig:performance}b shows an example of a scaled pressure distribution for a pool boiling problem that highlights the sharp jump in values that are effectively captured by our numerical implementation. The corresponding weak scaling study of CPU-only performance of the Poisson solver is shown in Figure \ref{fig:performance}c, which clearly demonstrates an order of magnitude improvement in performance with \amrex and its native multigrid solver in comparison to implementation with \paramesh + \hypre.

\section{Outflow Forcing} \label{sc:outflow-forcing}
\begin{figure}[h]
    \centering
    \includegraphics[width=0.75\textwidth]{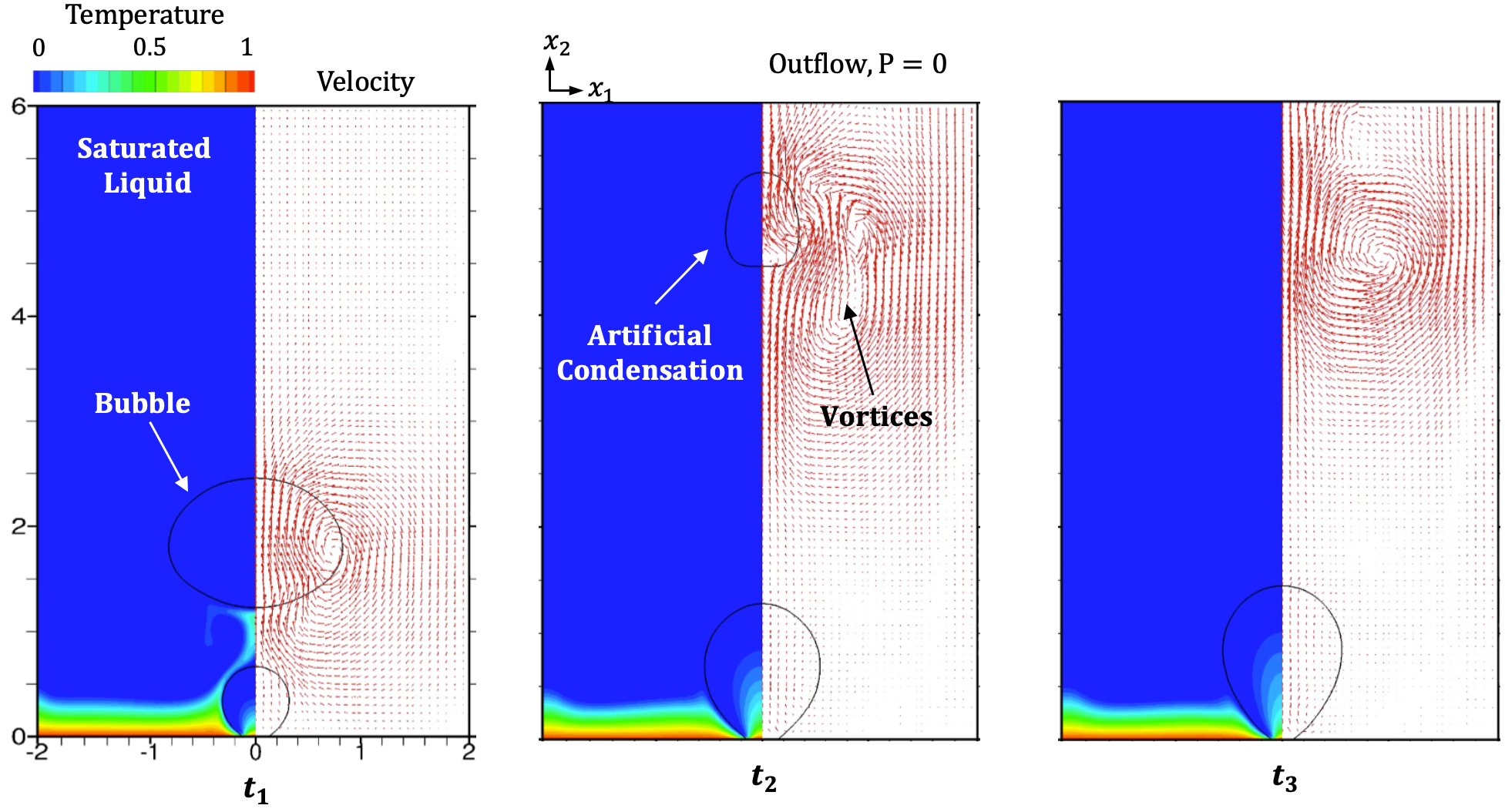}
    \caption{Effect of artificial condensation on saturated nucleate boiling. Physical condensation does no exist since both the bubble and bulk liquid are at the saturation temperature, however, outflow forcing in \cite{DHRUV2019, Sato2013} introduces artificial condensation which is non-physical ($t_1<t_2<t_3)$.}
    \label{fig:artificial_condensation}
\end{figure}

{The goal of proposed outflow forcing is to improve on previous implementations \cite{DHRUV2019,Sato2013} and provide a more concrete approach that avoids the need for introducing artificial condensation within the computational domain. Artificial condensation, as discussed previously, is a non-physical artifact which was implemented to ensure existence of only liquid phase at the outflow boundary. Figure \ref{fig:artificial_condensation} describes this behavior in time for a saturated nucleate boiling problem. Even though this approach helps with the multiphase instabilities at the boundary, introduction of artificial condensation affects the overall quality of the numerical simulation over time. In the present approach, we focus on relaxing the flow to a steady state instead by deriving forcing terms that have physical meaning within the context of multiphase flows.} 

\begin{figure}[h]
    \centering
    \includegraphics[width=0.7\textwidth]{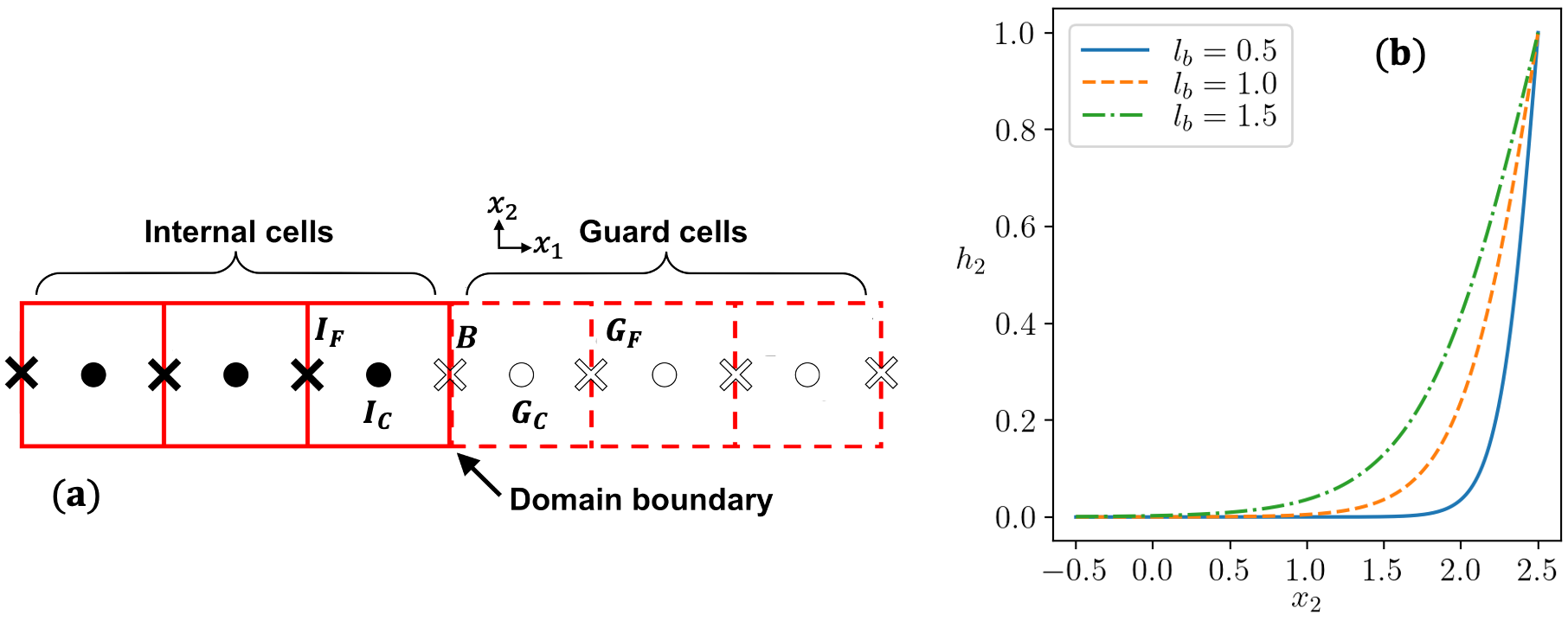}
    \caption{(a) Staggered computational grid near the outflow with faces ($\times$) and cell-centers ($\bullet$), $I$ denotes the internal cells, $G$ denotes the guard cells, and $B$ is the domain boundary. (b) Shape of outflow profiles, $h_2$, for different buffer region lengths, $l_b$.}
    \label{fig:stencil_outflow}
\end{figure}

Within our numerical formulation, vectors and scalars are spatially represented on a staggered grid. Vectors such as velocity are located on faces ($\times$), and scalars such as pressure are located at the cell centers ($\bullet$) as shown in the stencil in Figure \ref{fig:stencil_outflow}. The implementation of a staggered grid enables a tight coupling between the pressure gradient and the velocity during the corrector step in Equation \ref{eq:corrector} which translates into a more robust implementation of the Poisson equation for pressure in Equation \ref{eq:poisson}. { The predictor-corrector Equation \ref{eq:predictor} and \ref{eq:corrector} at the outflow have the following form,

\begin{equation} \label{eq:pred-corr-bnd-a}
    u^*_1 \Bigr|_{B} = u^n_1 \Bigr|_{B} + \Delta t \Bigg( \frac{3}{2}\Theta^n_1 - \frac{1}{2}\Theta^{n-1}_1 + \text{S}^{\Gamma}_{u_1} + \text{S}^{\text{O}}_{u_1} \Bigg)\:\Biggr|_{B} + \Delta t \frac{g_1}{\Fr^2}
\end{equation}
{\begin{equation} \label{eq:pred-corr-bnd-b}
\begin{split}
    u^{n+1}_1 \Bigr|_{B} = u^*_1 \Bigr|_{B} + \eqnmarkbox[blue]{book-keeping}{\Delta t \Bigg( \frac{1}{\rho_G'} (\text{S}^{\Gamma}_{\partial_1 P} + \text{S}^{\text{O}}_{\partial_1 P}) + \Bigg(\frac{1}{\rho_G'} - \frac{1}{\rho'} \Bigg) (2\zeta^n_1 - \zeta^{n-1}_1) \Bigg)\:\Biggr|_B} - \eqnmarkbox[teal]{ppe}{\Delta t \frac{1}{\rho_G'} \frac{P^{n+1}|_{G_C}-P^{n+1}|_{I_C}}{\Delta x_1}}
\end{split}
\end{equation}}
\annotate[yshift=-0.05em,color=black]{below,right}{ppe}{Solution from PPE}
\annotate[yshift=-0.05em,color=black]{below,left}{book-keeping}{Pressure correction and surface tension forces}

During the solution of Equation \ref{eq:pred-corr-bnd-a} and \ref{eq:pred-corr-bnd-b}, the pressure ($P^{n+1}$) at the outlet is set equal to $0$ by extrapolating values from $I_C$ to $G_c$,

\begin{equation}
    P^{n+1}|_{G_C} = - P^{n+1}|_{I_C}
\end{equation}

Consequently, $u^{n+1}_1$ at the domain boundary $B$ is computed using the pressure gradient during the corrector step in Equation \ref{eq:pred-corr-bnd-b}. The velocity values at the face-center guard-cells, $G_F$, are then extrapolated from $I_F$ to satisfy a zero gradient in the direction normal to the boundary, $u^{n+1}_1|_{G_F} = u^{n+1}_1|_{I_F}$. The solution of PPE corrects the predicted velocity, $u^*_1$ in Equation \ref{eq:pred-corr-bnd-b} and implicitly satisfies mass conservation, therefore the value of $u^{n+1}_1$ at domain boundary should remain untouched after the solution. For inlet, slip and no-slip boundary conditions, the pressure gradient at the boundary is set to zero which also enforces continuity for $u^{n+1}_1$ at the boundary $B$.

{The overall algorithm for the fractional-step method is described as follows,

\begin{enumerate}
    \item Solve level-set advection Equation \ref{eq:levelset-convection}, and compute variable properties described in Equation \ref{eq:properties} and \ref{eq:smeared-density}.
    \item Compute source terms $\text{S}^{\Gamma}_{u_i}$, $\text{S}^{\Gamma}_{\partial_i P}$, and $\text{S}^{\Gamma}_{T}$.
    \item Solve advection-diffusion Equation \ref{eq:temp} for temperature.
    \item Compute $\dot{m}$ using Equation \ref{eq:evaporation}
    \item Compute outflow forcing terms $\text{S}^{O}_{u_i}$, $\text{S}^{O}_{\partial_i P}$ described in Equation \ref{eq:vel-outflow} and \ref{eq:pres-outflow}.
    \item Solve Equation \ref{eq:predictor} for predicted velocity, $u^*_i$.
    \item Enforce boundary conditions for $u^*_i$, including at face-center boundary, $B$.
    \item Solve PPE Equation \ref{eq:poisson} using multigrid methods supplied by \amrex. Boundary conditions for $P^{n+1}$ are built into the linear system.
    \item Solve Equation \ref{eq:corrector} for corrected velocity, $u^{n+1}_i$.
    \item Enforce boundary conditions only $u^{n+1}_i$, except at face-center boundary, $B$.
\end{enumerate}
}}

The forcing terms for outflow $\text{S}^{\text{O}}_{u_i}$ and $\text{S}^{\text{O}}_{\partial_i P}$ are applied to internal cells within a buffer region of length, $l_b$, near the outflow boundary. To define the outflow region we use a weighting expression,

\begin{equation} \label{eq:outflow-profile}
    h_i = \Bigg(\frac{2}{1+e^{4({x^{max}_i-x_i})/l_b}}\Bigg)
\end{equation}

where, $x$ is the spatial location of the grid point, $x^{max}$ is the upper bound of the domain, $i=\{1,2,3\}$ is the tensor notation for the spatial directions. Fig \ref{fig:stencil_outflow}b shows the profile of this function for different values of $l_b$. The choice of this profile was motivated by the need to concentrate the effect of the forcing function only near the boundary while also avoid conditional statements within stenciled computation to enable loop unrolling and optimization. The exponential function in Equation \ref{eq:outflow-profile} served as a suitable candidate for this requirement. {Choice for $l_b$ is governed by the size of disturbances that exit out of the computational domain. For multiphase flows like boiling described in Fig \ref{fig:schematic} these disturbances are proportional to the capillary length, $l_c = \sqrt{\sigma/(\rho_L-\rho_G) \: g}$, of the liquid-gas pair and the width of the channel in direction parallel to boundary. As we demonstrate in Section \ref{sc:results}, choosing $l_b$ within this range adequately damps fluctuations that cause instabilities in the flow and preserves statistical behavior of the solution in the upstream region.}

The phase averaged flow within the buffer region is given as the integral over control volume of the domain, $V$,

\begin{equation} \label{eq:outflow-mean}
\overline{u}_{L_i} = \sum_{j=1}^{3} \delta_{ij} \frac{\int_V (1-\text{H}) (u_j^n h_j \hat{o}_j) dV}{\int_V (1-\text{H}) (h_j \hat{o}_j) dV},\:\: \overline{u}_{G_i} = \sum_{j=1}^{3} \delta_{ij} \frac{\int_V \text{H} (u_j^n h_j \hat{o}_j) dV}{\int_V \text{H} (h_j \hat{o}_j) dV}
\end{equation}

where, $\hat{o}_j = 1$, if the flow is defined in the direction $j$, or $0$ otherwise. $\delta_{ij}$ is the Kronecker delta, and $\text{H}$ is the sharp Heaviside function from Equation \ref{eq:heaviside}a, 

\begin{equation} \label{eq:kronecker}
\delta_{ij} = \begin{cases}
     1,& \text{if} \: i = j \\
     0,& \text{if} \: i \neq j\\
\end{cases} 
\end{equation}

The effective outflow velocity, $u_E$, is given by,
\begin{equation} \label{eq:outflow-eff}
u_{E_i} = \text{max} \big({u_c}, \overline{u}_{L_i}  \big) (1-\text{H}) + \text{max} \big({u_c}, \overline{u}_{G_i} \big) (\text{H})
\end{equation}

setting the lower bound to terminal velocity, ${u_c = \sqrt{g l_c}}$. The outflow forcing terms are subsequently given by, 

\begin{equation} \label{eq:vel-outflow}
\begin{split}
\text{S}^{\text{O}}_{u_i} = \eqnmarkbox[red]{stream}{\sum_{j=1}^{3} \delta_{ij} h_j \hat{o}_j \vast[\frac{\text{min} \Big(1, \big\lvert\frac{u_{E_j}}{u_j^n}\big\rvert\Big) u_j^n - u_j^n}{\Delta t}\vast]} + \eqnmarkbox[teal]{normal} {\sum_{j=1}^{3} \sum_{k=1}^{3} \delta_{ij} (1- \delta_{jk}) h_k \hat{o}_k \vast[\frac{\text{min} \Big(1, \big\lvert\frac{u_{E_k}}{u_j^n}\big\rvert\Big) u_j^n - u_j^n}{\Delta t}\vast]} \\
- \eqnmarkbox[blue]{convection} {\sum_{j=1}^{3} h_j \hat{o}_j \lvert u_{E_j} \rvert \partial_j u_i^n}
\end{split}
\end{equation}
\annotate[yshift=-0.05em,color=black]{below,right}{stream}{Scaling in streamwise direction}
\annotate[yshift=0.75em,color=black]{above,right}{normal}{Scaling in normal direction}
\annotate[yshift=-0.25em,color=black]{below,right}{convection}{convection}

\begin{equation} \label{eq:pres-outflow}
\text{S}^{\text{O}}_{\partial_i P} = - \eqnmarkbox[red]{stream}{\sum_{j=1}^{3} \delta_{ij} h_j \hat{o}_j \text{S}^{\Gamma}_{\partial_i P}} - \eqnmarkbox[teal]{normal} {\sum_{j=1}^{3} \sum_{k=1}^{3} \delta_{ij} (1- \delta_{jk}) h_k \hat{o}_k \text{S}^{\Gamma}_{\partial_j P}}
\end{equation}
\annotate[yshift=0.5em,color=black]{above,right}{stream}{Forcing in streamwise direction}
\annotate[yshift=-0.25em,color=black]{below,left}{normal}{Forcing in normal direction}

{Each term in Equations \ref{eq:vel-outflow} and \ref{eq:pres-outflow} is highlighted to show its significance. The buffer region represents a far-field with a mean velocity, $u_e$. Local velocities with magnitudes greater than $u_e$ are scaled in both streamwise and normal direction. It is important to note that scaling does not alter the direction of the velocity and only ensures that magnitude of local velocities cannot be greater than $u_e$. Natural backflow can still exist with $P=0$ outflow boundary condition. Enforcing the condition $u_e \geq u_c$ ensures, at the very least, that capillary flow conditions are satisfied within the buffer region. The convection term adds another forcing to allow turbulent vortices to exit the domain. $\text{S}^{\text{O}}_{\partial_i P}$ counteracts pressure jumps, $\text{S}^{\Gamma}_{\partial_i P}$, and completely nullifies its contribution at the outflow, which aligns with the far-field assumption within the buffer region.} In the following section we present results from some numerical experiments that demonstrate efficacy of this outflow forcing.
%
%
\section {Results} \label{sc:results}
This section presents a set of verification problems to demonstrate the stability of outflow forcing described in the section above. Problems are presented with increasing complexity, starting from benchmark multiphase flow problems to homogeneous evaporation and heterogeneous pool/flow boiling. We use laboratory notebooks that organizes each computational experiment using a set of configuration files to enable data curation and reproducibility\cite{dhruv_dubey_2023}. The laboratory notebook and Flash-X source code are open source to allow community development and contribution \cite{outflow_forcing_bubbleml,outflow_forcing_revised,flow_boiling_performance}.

\subsection{Comparison with multiphase flow benchmarks} \label{sc:benchmarking}

{We start with a benchmark problem of a two-dimensional rising bubble without phase transition. We chose this problem because of availability of data from multiple solvers \cite{bubble_benchmarks}. Figure \ref{fig:rising_bubble_schematic}(a) describes the computational setup used in developing these benchmarks, and Figure \ref{fig:rising_bubble_schematic}(b) describes the setup that replaces the no-slip boundary condition at the top with a pressure outflow and buffer regions to allow for testing our formulation. We perform simulations using both configurations to allow for comparing results from different \flashx implementations and other solvers.
\begin{figure}[ht]
    \centering
    \includegraphics[width=0.5\textwidth]{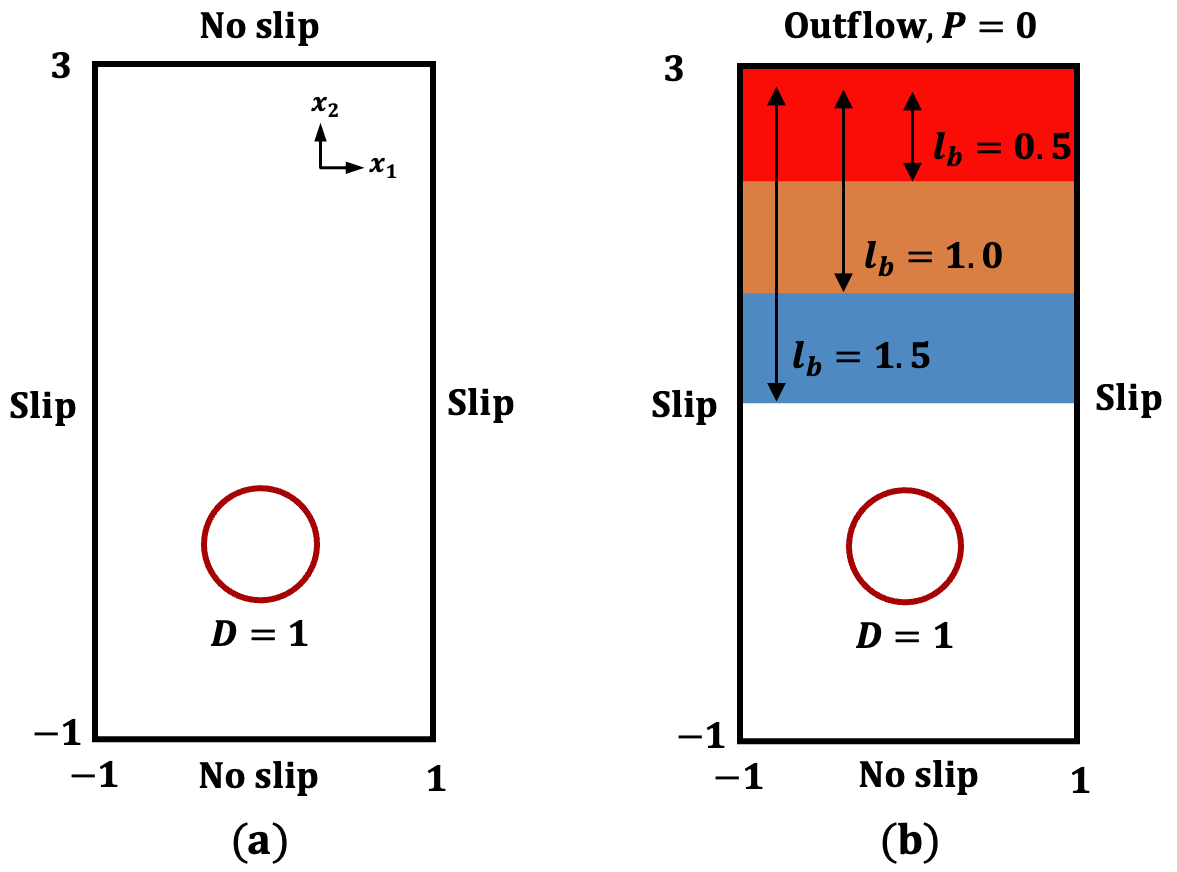}
    \caption{Schematic of the benchmark rising bubble problem (a) With a no-slip boundary condition at the top of the domain, and (b) With outflow boundary condition along with buffer regions for testing}
    \label{fig:rising_bubble_schematic}
\end{figure}

Flow configurations provided in \cite{bubble_benchmarks} allowed us to calculate non-dimensional parameters described in Table \ref{table:rising}, along with the reference length scale ($l_0=0.5 m$), and time scale ($t_0=0.7s$). $l_0$ corresponds to the diameter of the bubble. Simulations are initialized by placing the bubble with $D=1$ at the center $(0,0)$ of the computational domain, which is also scaled using $l_0$.
\begin{table}[h]
\begin{center}
\begin{tabular}{|c|c|c|c|c|c|}
\hline
Parameter & $ \rho_G/\rho_L$ & $\mu_G/\mu_L$ & $\text{Re}$ & $\text{Fr}$ & $\text{We}$ \\
\hline
Values  & $0.001$ & $0.01$ & $35$ & $1$ & $125$  \\ \hline
\end{tabular}
\caption{Parameters for the rising bubble benchmark problem \label{table:rising}}
\end{center}
\end{table}

Starting with a resolution of $D/80$, we simulate all the configurations described in Figure \ref{fig:rising_bubble_schematic}. To compare simulation results, we track evolution of circularity ($C$), bubble center of mass ($X_2$), and bubble rise velocity ($U_2$) with time. Equations for these quantities are described in \cite{bubble_benchmarks} and elaborated below,

\begin{enumerate}
    \item $C$: Ratio of area-equivalent circle to perimeter of the bubble. For a perfect circle, $C=1$, which decreases as the bubble deforms.
    \item $X_2$: Normal component of the bubble center of mass, computed using integral of area occupied by the bubble.
    \item $U_2$: Normal component of the mean velocity of the bubble.
\end{enumerate}
\begin{figure}[ht]
    \centering
    \includegraphics[width=0.9\textwidth]{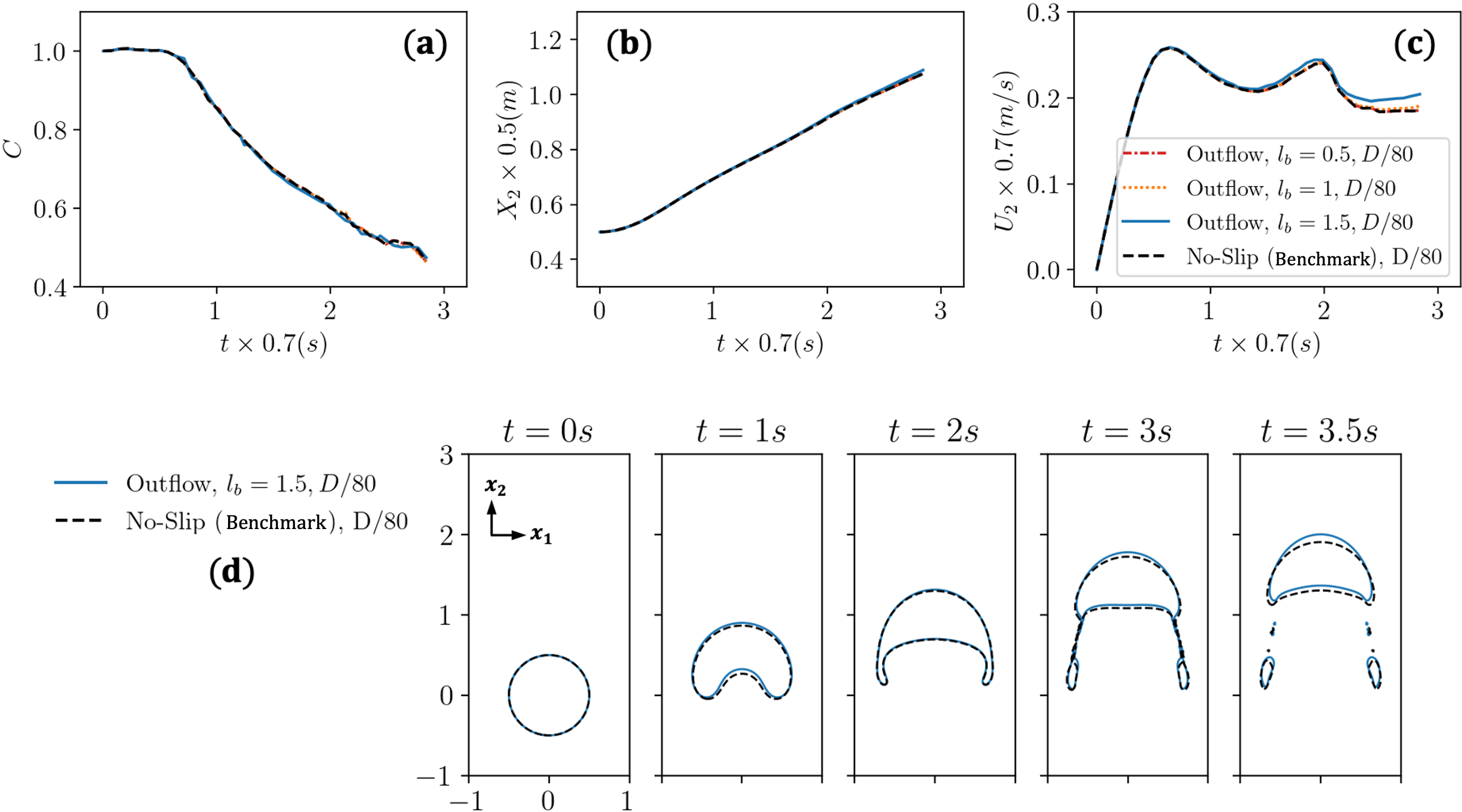}
    \caption{Comparison of benchmark quantities for different simulation configurations. {The no-slip case refers to the setup reported in \cite{bubble_benchmarks}, which is compared with the setup that uses the outflow forcing}.}
    \label{fig:rising_bubble_outflow}
\end{figure}

{From the results we can see that the outflow forcing closes matches the no-slip solution for $l_b=\{0.5,1\}$, and starts to deviate when $l_b>D$, the size of the disturbance in this case}. The deviation is particularly prominent in the evolution of rise velocity, $U_2$ (Figure \ref{fig:rising_bubble_outflow}c), which is a direct consequence of discrepancy in the bubble break-up pattern at $t=3.5s$ (Figure \ref{fig:rising_bubble_outflow}d). Sensitivity of bubble break-up event for this benchmark problem has been reported in previous studies \cite{bubble_benchmarks} as well, with different numerical formulations exhibiting different behaviors. 
\begin{figure}[ht]
    \centering
    \includegraphics[width=0.9\textwidth]{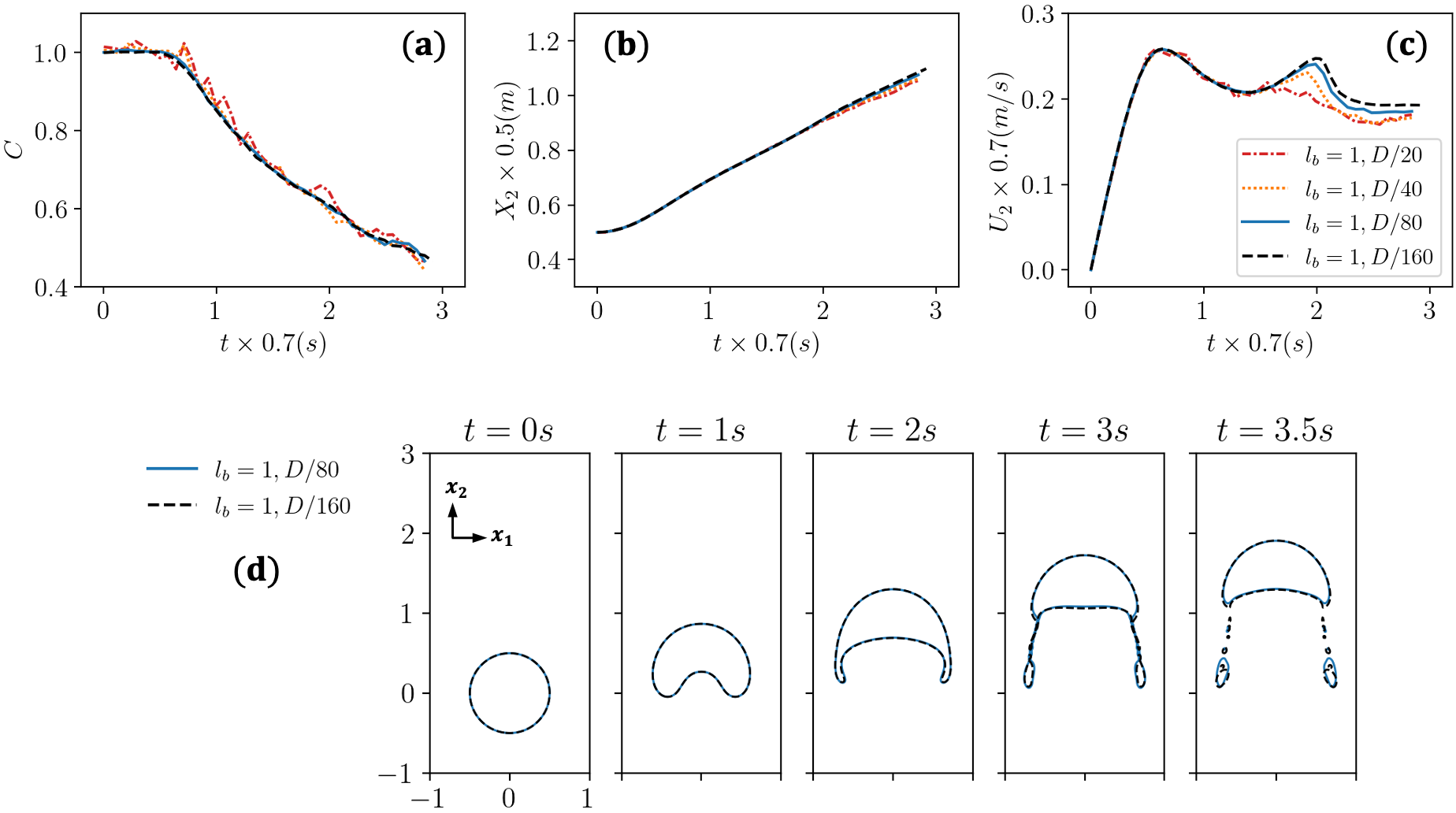}
    \caption{Comparison of benchmark quantities for different spatial resolutions. Coarser grids, $D/20$ and $D/40$ demonstrate noisy behavior while $D/80$ and $D/160$ show good convergence, although minor discrepancy can still be observed in the bubble break-up behavior. Due to the cost associated with simulating this problem we could perform a study for $D/320$.}
    \label{fig:rising_bubble_convergence}
\end{figure}
\begin{table}
\begin{center}
\begin{tabular}{|c|c|c|c|c|c|c|c|}
\hline
$\Delta x_i$ & $\Delta t$ & $||e_1||$ & $\text{ROC}_1$ & $||e_2||$ & $\text{ROC}_2$ & $||e_{\infty}||$ & $\text{ROC}_\infty$\\
\hline
$$D/20$$ & $5.0E-05$ & $2.9E-02$ & - & $3.8E-02$ &  & $6.4E-02$ & - \\
\hline
$$D/40$$ & $1.25E-05$ & $1.4E-02$ & 1.1 & $1.9E-02$ & 1.0 & $5.8E-02$ & -\\
\hline
$$D/80$$ & $3.125E-06$ & $5.8E-03$ & 1.3 & $7.2E-03$ & 1.4 & $1.3E-02$ & 2.2 \\
\hline
\end{tabular}
\caption{Relative norms and convergence rate of bubble circularity, $C$. \label{table:rising-bubble-circularity}}
\end{center}
\end{table}
\begin{table}
\begin{center}
\begin{tabular}{|c|c|c|c|c|c|c|c|}
\hline
$\Delta x_i$ & $\Delta t$ & $||e_1||$ & $\text{ROC}_1$ & $||e_2||$ & $\text{ROC}_2$ & $||e_{\infty}||$ & $\text{ROC}_\infty$\\
\hline
$$D/20$$ & $5.0E-05$ &$9.8E-03$ & - & $1.9E-02$ &  & $3.5E-02$ & -\\
\hline
$$D/40$$ & $1.25E-05$ &$5.7E-03$ & 0.8 & $1.0E-02$ & 0.9 & $1.9E-02$ & 0.88\\
\hline
$$D/80$$ & $3.125E-06$ & $2.1E-03$ & 1.44 & $3.8E-03$ & 1.39 & $8.5E-03$ & 1.16 \\
\hline
\end{tabular}
\caption{Relative norms and convergence rate of bubble center of mass, $X_2$. \label{table:rising-bubble-center}}
\end{center}
\end{table}
\begin{table}
\begin{center}
\begin{tabular}{|c|c|c|c|c|c|c|c|}
\hline
$\Delta x_i$ & $\Delta t$ & $||e_1||$ & $\text{ROC}_1$ & $||e_2||$ & $\text{ROC}_2$ & $||e_{\infty}||$ & $\text{ROC}_\infty$\\
\hline
$$D/20$$ & $5.0E-05$ & $6.5E-02$ & - & $8.8E-02$ & - & $2.5E-01$ & - \\
\hline
$$D/40$$ & $1.25E-05$ & $4.0E-02$ & 0.7 & $6.3E-02$ & - & $1.6E-01$ & 0.64 \\
\hline
$$D/80$$ & $3.125E-06$ & $1.5E-02$ & 1.4 & $2.5E-02$ & 1.33 & $6.0E-02$ & 1.4 \\
\hline
\end{tabular}
\caption{Relative norms and convergence rate of bubble velocity, $U_2$. \label{table:rising-bubble-velocity}}
\end{center}
\end{table}

Next we perform a grid refinement study using the outflow forcing configuration with $l_b=1$ to estimate the convergence rate of our numerical formulation. The results for this study are reported in Figure \ref{fig:rising_bubble_convergence}. Coarser grids, $D/20$ and $D/40$, demonstrate a lot of noise in the evolution of circularity ($C$) and rise velocity ($U_2$). Even though good convergence is achieved between $D/80$ and $D/160$, the solutions are far from grid independent, which is clearly evident in the evolution of the rise velocity ($U_2$) which is again a result of discrepancy in the bubble break-up behavior. We could not perform a simulation at resolution of $D/320$ due to the time-stepping constraints imposed by the explicit nature of our formulation.

Tables \ref{table:rising-bubble-circularity}, \ref{table:rising-bubble-center}, and \ref{table:rising-bubble-velocity}, provide error norms and convergence rates for the spatial resolution study in Figure \ref{fig:rising_bubble_outflow}. The norms are computed using $D/160$ as the reference solution. Convergence rates are computed using,

\begin{equation}
    ROC = \frac{log(||e||_{coarse}/||e||_{fine})}{log(\Delta x_{coarse}/\Delta x_{fine})}
\end{equation}

Due to the noisy behavior on coarser grids, the convergence rates between $D/20$ and $D/40$ the convergence rates between them are less than linear. However rates between $D/40$ and $D/80$ are greater than linear nearing a second-order convergence.

Finally we compare our simulations results from the finest resolution to the studies reported previously \cite{bubble_benchmarks} (Figure \ref{fig:rising_bubble_benchmarks}. Perfect agreement between results is not obtained due to the sensitivity of the study. However, general trend in circularity, $C$, and velocity ($U_2$) is captured. Notice the timing of two peaks for $U_2$. The first is peak observed when bubble reaches terminal velocity, and the second is observed near the breakup. 
\begin{figure}[ht]
    \centering
    \includegraphics[width=0.9\textwidth]{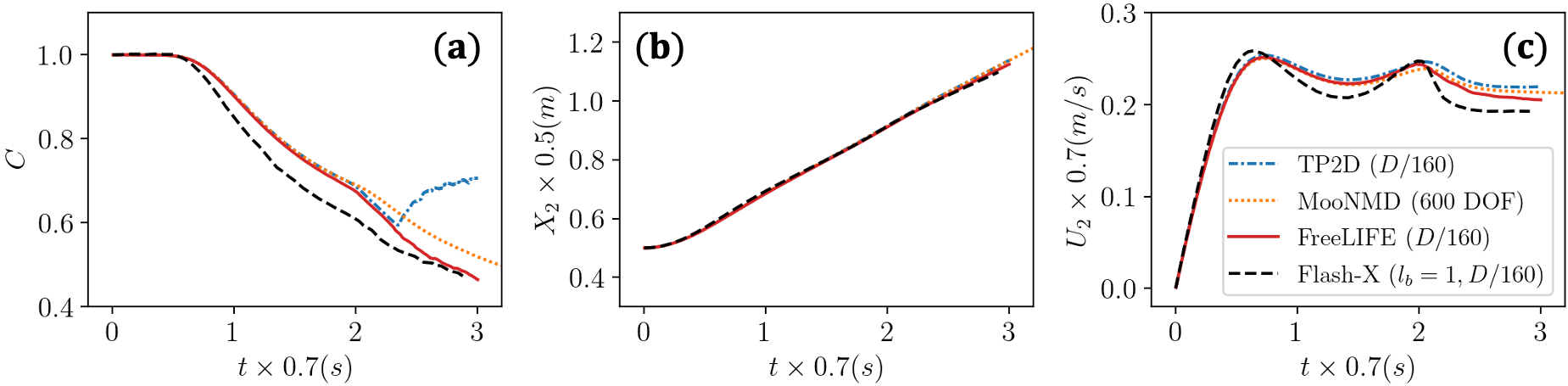}
    \caption{Comparison of \flashx simulation results for rising bubble benchmark with previously reported studies in \cite{bubble_benchmarks}. Due to the sensitivity of the problem the agreement is not perfect between any of the dataset, however the general trend in circularity ($C$) and velocity ($U_2$) is captured.}
    \label{fig:rising_bubble_benchmarks}
\end{figure}
}
\subsection{Bubble evaporation in superheated liquid} \label{sc:bubble-evap}

\begin{figure}[ht]
    \centering
    \includegraphics[width=0.25\textwidth]{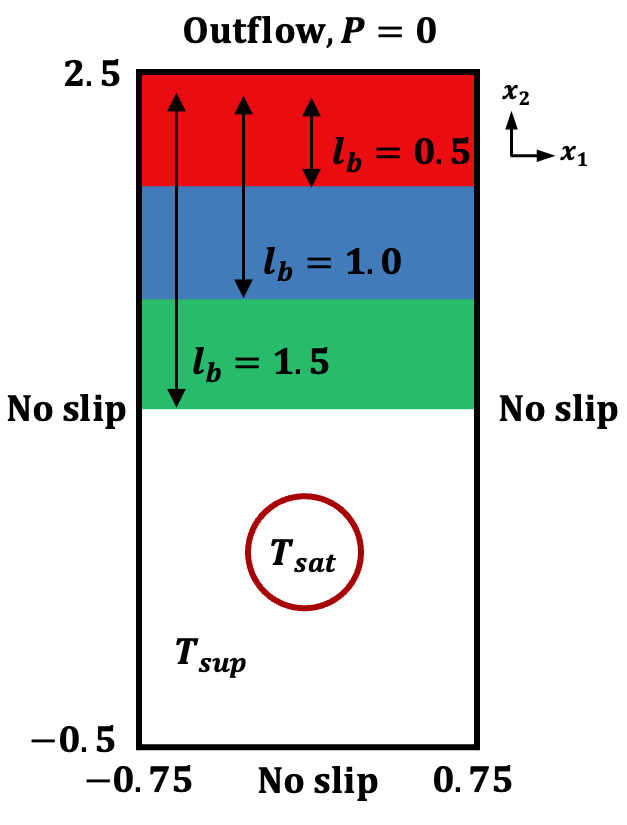}
    \caption{Schematic of the evaporating bubble problem with corresponding outflow profiles, $h_2$, for different buffer region lengths, $l_b$.}
    \label{fig:evaporating_bubble_schematic}
\end{figure}

{Two dimensional homogeneous evaporation of a gas (vapor) bubble in superheated liquid is another case to compare the effect of the outflow forcing with a solution where no forcing is applied ($\text{S}^{\text{O}}_{u_i}=\text{S}^{\text{O}}_{\partial_i P}=0$)}. The schematic of the problem is provided in Figure \ref{fig:evaporating_bubble_schematic} with dimensions of the domain in directions $x_1$ and $x_2$. The simulation is initialized by placing a spherical bubble of radius $r=0.25$ and center at the coordinates $(0,0)$. The temperature inside the bubble is set to $T_{sat}=0$, and the temperature inside the liquid is set to $T_{sup}=1$. Table \ref{table:evaporating} provides values for the remaining non-dimensional parameters for the simulation. {A resolution of $D/16$ along with a time step of $2.5E-05$ is used for this study.}
\begin{table}
\begin{center}
\begin{tabular}{|c|c|c|c|c|c|c|c|c|c|c|}
\hline
Parameter & $ \rho_G/\rho_L$ & $\mu_G/\mu_L$ & $C_{p_G}/C_{p_L}$ & $k_G/k_L$ & $\alpha_G/\alpha_L$ & $\text{Re}$ & $\text{Pr}$ & $\text{St}$ & $\text{Fr}$ & $\text{We}$ \\
\hline
Values       & $0.0012$ & $0.00025$ & $0.7$ & $0.2$ & $238$ & $200$ & $7$ & $0.004$ &  $1$ & $50$ \\ \hline
\end{tabular}
\caption{Parameters for evaporating bubble case \label{table:evaporating}}
\end{center}
\end{table}

No-slip boundary conditions are applied everywhere, except for the outflow where $P=0$. No slip boundary conditions set the velocities and normal derivative of pressure at the boundaries to zero. The effect of the outflow forcing is tested for different buffer lengths, $l_b = \{0.5, 1.0, 1.5\}$, and compared with the no forcing solution.
\begin{figure}[h]
    \centering
    \includegraphics[width=0.95\textwidth]{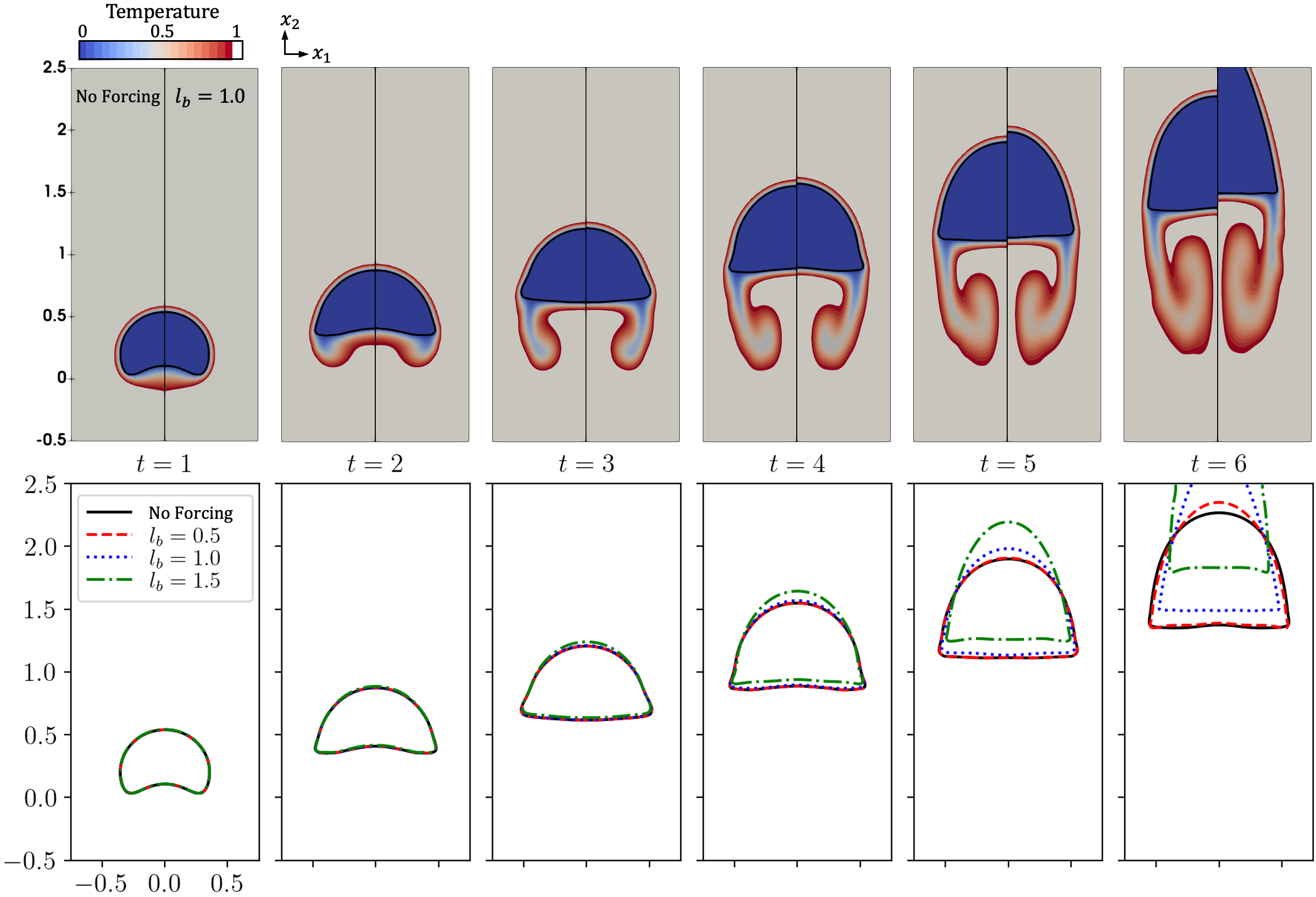}
    \caption{Comparison of temperature (top), and evolution of liquid-gas interface, $\phi=0$ (bottom), between no forcing solution and solution obtained using outflow forcing for varying $l_b$.}
    \label{fig:evaporating_bubble_compare}
\end{figure}
\begin{figure}[h]
    \centering
    \includegraphics[width=0.95\textwidth]{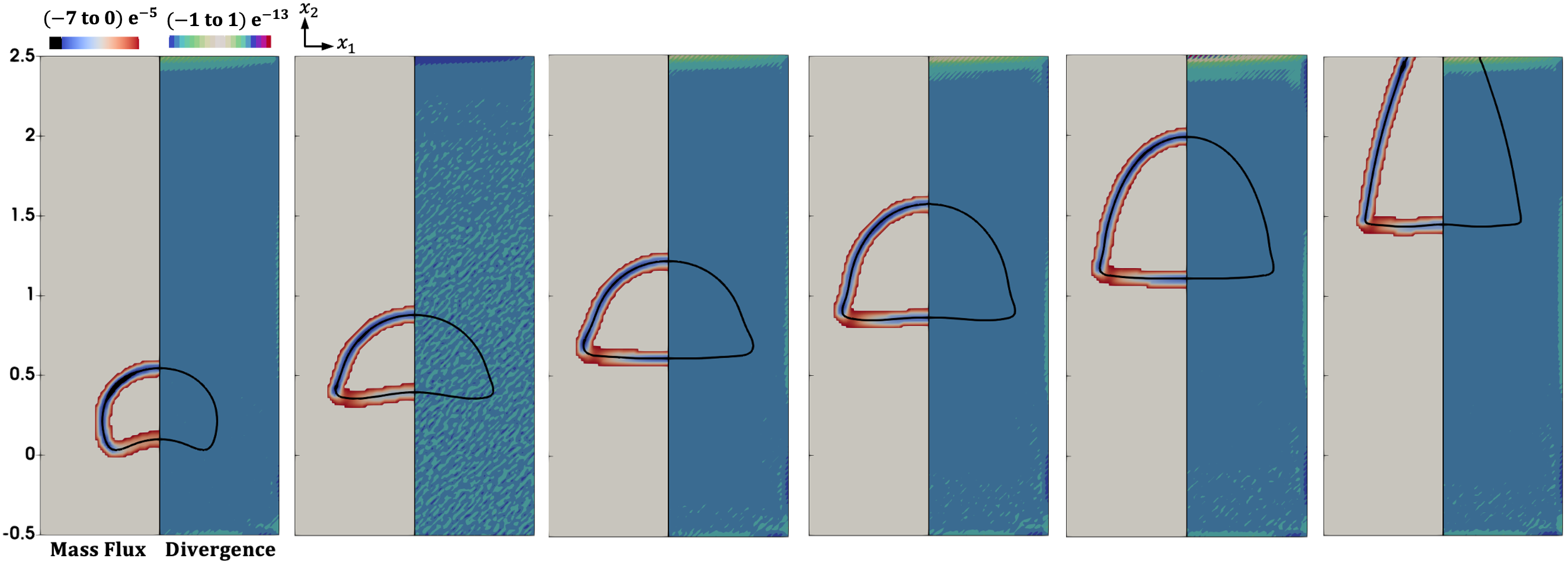}
    \caption{ Evolution of evaporative mass flux, $\dot{m}$ (left), and divergence of velocity, $\sum_{j=1}^{3}\partial_j u_j + \sum_{j=1}^{3} \dot m \: n_{\Gamma_j} \partial_j  \frac{1}{\rho'} \:\bigr\vert_{\:\Gamma}$ (right) demonstrating the divergence preserving nature of the outflow forcing ($l_b=1.0$).}
    \label{fig:evaporating_bubble_div}
\end{figure}

{Figure \ref{fig:evaporating_bubble_compare} provides the temporal evolution for different cases, which show that the effects of the outflow forcing only appear in the buffer region and maintain the accuracy of the flow in the downstream areas. The temperature contours and bubble shape are consistent for $x^{max}_i - x_i > l_b$. For $x^{max}_i - x_i \leq l_b$, the bubble undergoes acceleration due to the presence of the convective term in Equation \ref{eq:vel-outflow}, which changes the shape of the liquid-gas interface and the corresponding temperature distribution. This test problem provides good verification of the sharp nature of the outflow forcing.}

{Divergence preserving nature of our formulation is evident in Figure \ref{fig:evaporating_bubble_div} where temporal evolution of divergence is of order ${10E-13}$. The figure also show evolution of evaporative mass flux, $\dot{m}$, which exists only near the liquid-gas interface.}

\subsection{Two dimensional pool boiling simulation with a single nucleation site}

\begin{figure}[h]
    \centering
    \includegraphics[width=0.95\textwidth]{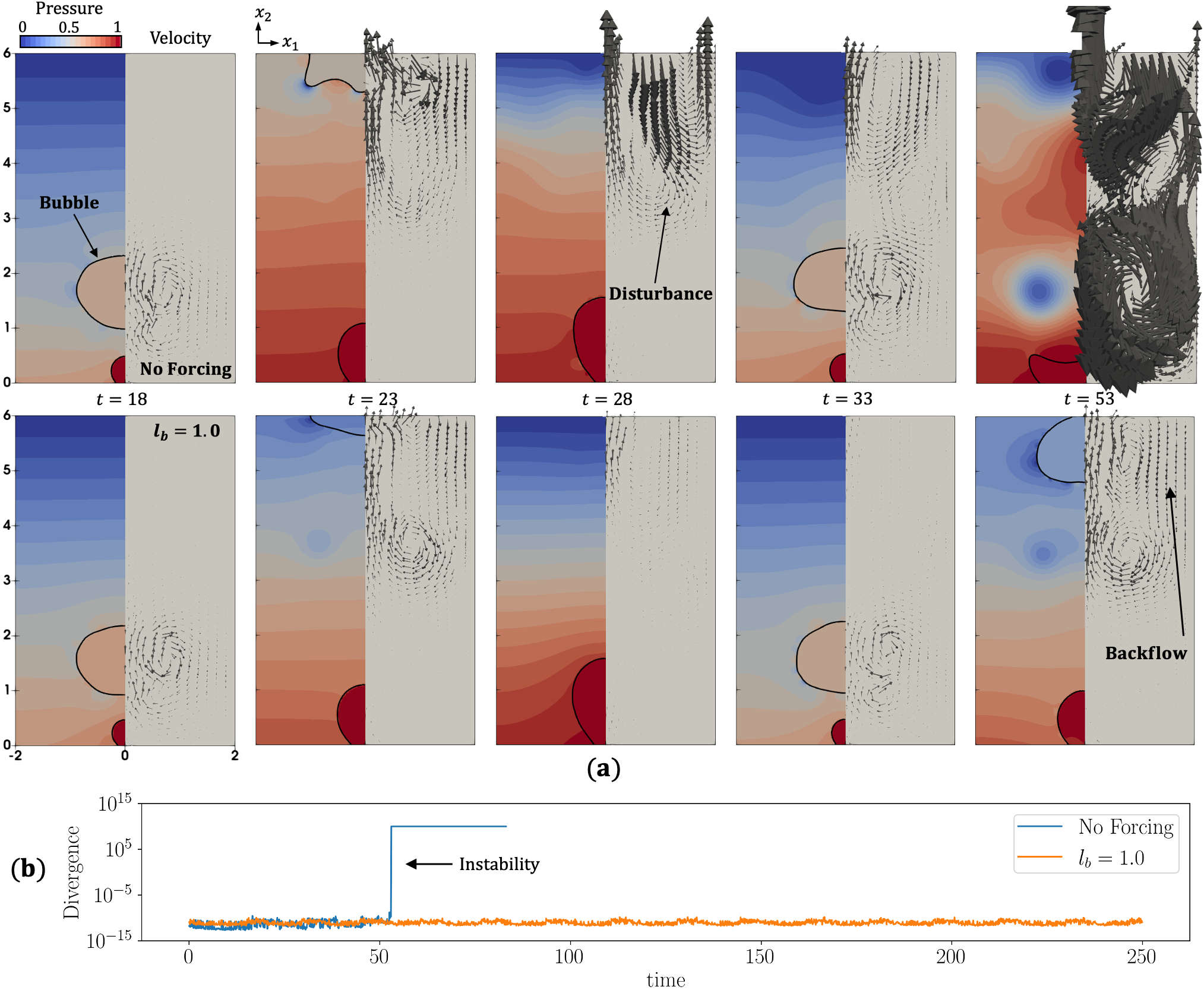}
    \caption{(a) Comparison of pressure and velocity during the intial cycles of bubble growth and departure for simulation without outflow forcing (top) and simulation with $l_b=1.0$ (bottom). (b) Evolution of divergence with time.}
    \label{fig:pool_boiling_velocity}
\end{figure}

The next problem we consider is the simplified version of the pool boiling problem described in Figure \ref{fig:schematic}b, where we look at dynamics associated with a single nucleation site. This simulation not only adds an additional layer complexity due to the heterogeneous nature of the boiling process (boiling over a heater), but also provides an opportunity to understand the nature of vortex-induced instability at the outflow by considering effects associated with a single bubble. The size of the computational domain is shown in Figure \ref{fig:pool_boiling_velocity}, with the heater completely occupying the lower boundary in the direction $x_2$.

We follow the nucleation site model discussed in our previous work \cite{DHRUV2019,DHRUV2021,akash_phd_2021}, which provides site location and nucleation wait time, $t_{wait}$, as input to our simulation. The nucleation site model seeds a bubble of radius $0.1$ at a given nucleation site if it has been vacant (covered by liquid) for $t=t_{wait}$. For the test problem with a single nucleation site, the bubble is seeded at location $(0,0)$.

The dynamics of the contact line is modeled by satisfying a contact angle, $\psi$, for the level-set function, $\partial_2 \phi=-cos(\psi)$, at the heater boundary. $\psi$ is calculated using,
\begin{equation} \label{eq:contact-angle-eq}
\psi = \begin{cases}
    \psi_r,& \text{if } u_{base}< 0\\
    \frac{\psi_a-\psi_r}{u_{lim}}u_{base} + \psi_r,  & \text{if }  0 \leq u_{base} \leq u_{lim}\\
     \psi_a,& \text{if } u_{base}> u_{lim}\\
\end{cases}
\end{equation}
where, $\psi_r$ is the receding and $\psi_a$ is the advancing contact angle, $u_{base}$ is the near wall radial velocity around the interface in the plane parallel to heater and $u_{lim}$ is the limiting velocity. This model is a variation of that proposed by Mukherjee { et al.} ~\cite{Mukherjee2007}. They assume inertial effects to be relevant both in the advancing and receding conditions, whereas our model accounts for inertial effects only in the advancing state. From our numerical experiments and based on results from Mukherjee { et al.} we found that an optimum value for $u_{lim}$ is $20 - 25 \%$ of the characteristic velocity, $u_0 = \sqrt{g\:l_0}$, where $g$ is acceleration due to gravity, and $l_0$ is the characteristic length scale given by $l_0 =\sqrt{\sigma/(\rho_l-\rho_v) \: g}$. Table \ref{table:single-bubble} provides the values of the remaining input parameters for this simulation.
\begin{table}[h]
\begin{center}
\begin{tabular}{|c|c|c|c|c|c|c|c|c|c|c|c|c|c|c|}
\hline
Parameter & $ \rho_G/\rho_L$ & $\mu_G/\mu_L$ & $C_{p_G}/C_{p_L}$ & $k_G/k_L$ & $\alpha_G/\alpha_L$ & $\text{Re}$ & $\text{Pr}$ & $\text{St}$ & $\text{Fr}$ & $\text{We}$ & $\psi_r$ & $\psi_a$ & $t_{wait}$ & $T_{sat}$ \\
\hline
Values       & $0.0049$ & $0.02$ & $0.735$ & $0.143$ & $39$ & $303$ & $7$ & $0.16$ &  $1$ & $1$ & $45\degree$ & $45\degree$ & 0.4 & 0 \\ \hline
\end{tabular}
\caption{Parameters for pool boiling with single nucleation site \label{table:single-bubble}}
\end{center}
\end{table}

\begin{figure}[h]
    \centering
    \includegraphics[width=0.85\textwidth]{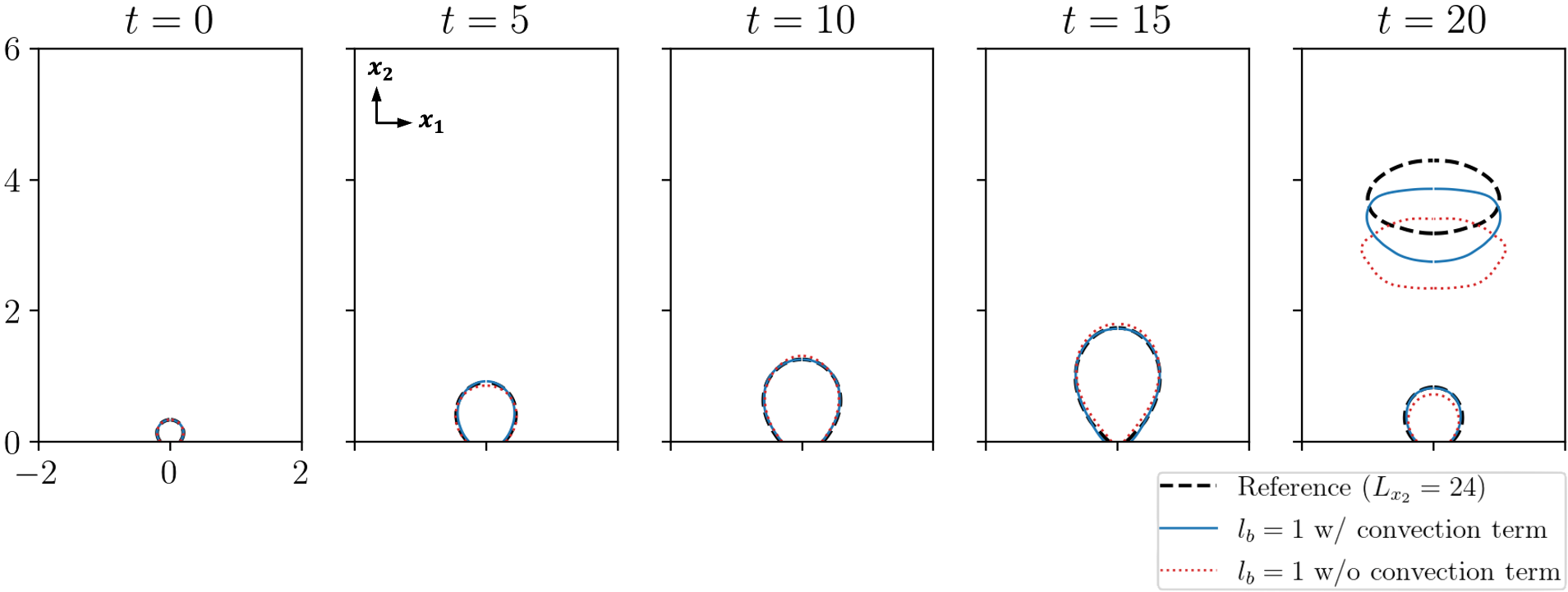}
    \caption{Effect of convection term in Equation \ref{eq:vel-outflow} term on the solution for pool boiling problem.}
    \label{fig:pool_boiling_advection}
\end{figure}

{Five different sets of studies were performed to compare the reference solution ($L_{x_2} = 24$), no forcing solution ($L_{x_2} = 6$), and three outflow forcing solution ($L_{x_2} = 6$) with different values of $l_b$. All simulations were performed using a fixed time step, $\Delta t = 10^{-4}$, and resolution, $\Delta x_i = 0.02$}. The comparison in Figure \ref{fig:pool_boiling_velocity}a for the initial cycles of bubble growth and departure shows the instabilities created at the exit when forcing is not applied. Each cycle consists of a single bubble nucleating, growing, and departing from the heater surface. The disturbances are the result of sharp interfacial jumps in pressure associated with $\text{We}=1$. The simulation between the two cases starts to deviate at $t = 53$, when the disturbances in the reference solution travel downstream and create unrealistic flow patterns that deform the nucleation of the bubble in the heater for the third cycle of bubble growth and departure. { Evolution of divergence in Figure \ref{fig:pool_boiling_advection}b, shows effect of this instability on growth of divergence in the domain which eventually causes the simulation to blow-up. By setting the outflow forcing within, $l_b=1.0$, quasi-steady state can be reached.} 

\begin{figure}[h]
    \centering
    \includegraphics[width=0.85\textwidth]{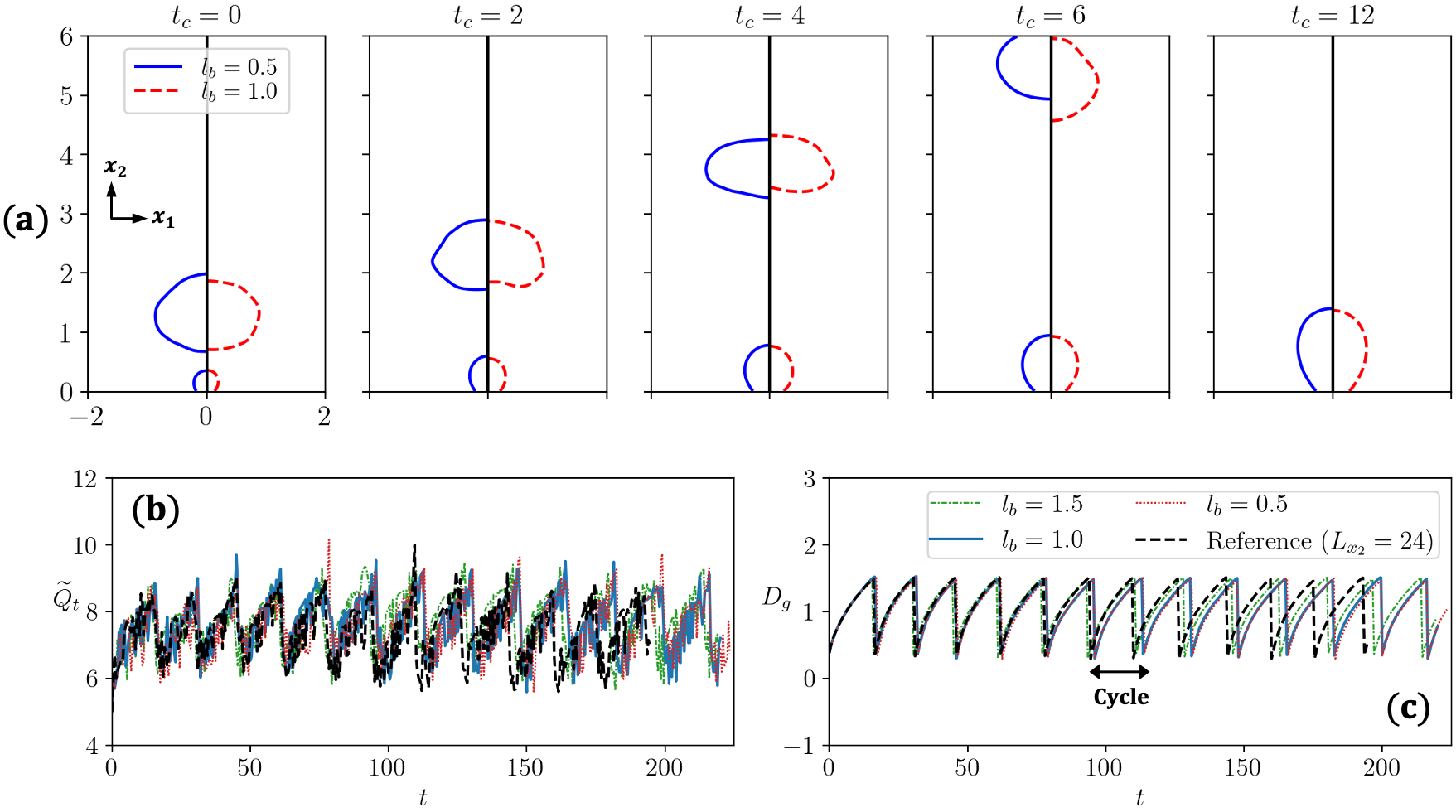}
    \caption{(a) Quasi-steady behavior of bubble growth and departure cycle for different values of $l_b$, (b) Temporal variation in the heat flux $\tilde{Q}$, and (c) Growth of equivalent bubble diameter $D_g$.}
    \label{fig:pool_boiling_compare}
\end{figure}

{The temporal and quasi-steady behavior shown in Figure\ref{fig:pool_boiling_compare} compares results obtained from different values of $l_b$ against each other and the reference solution ($L_{x_2}=24$). See the progression of the bubble growth and departure cycle from $t_c = 0-12$ in Figure\ref{fig:pool_boiling_compare}a. A more quantified comparison is found in Figures \ref{fig:pool_boiling_compare}b and c, which compares the behavior of the mean wall heat flux, $\tilde{Q}$, and the equivalent diameter during bubble growth, $D_g$ as a function of time. $\tilde{Q}$ is an integral value computed using,
\begin{equation} \label{eq:integral-heatflux}
    \tilde{Q} = \frac{\int_{S_{heater}} I_L \partial_2 T dS}{\int_{S_{heater}} dS}
\end{equation}
where, $S_{heater}$, is the surface area of the heater, $I_L$ is the indicator function for the liquid, and $\partial_2 T$ is the temperature gradient in the cell closest to the heater wall. 

The equivalent diameter, $D_g$, is computed using,
\begin{equation} \label{eq:eq-diam}
    D_g = 2 \sqrt{\frac{A_{bubble}}{\pi}}
\end{equation}
where, $A_{bubble}$, is the bubble area during the growth period. The temporal variation for both $\tilde{Q}$ and $D_g$ is consistent between the two runs with negligible differences.

It is evident that the outflow forcing introduces a phase difference in computation of these quantities, which is an expected effect of this formulation. However, the temporal trend is statistically consistent which is an important result. The value for $\overline{Q}$, 
\begin{equation} \label{eq:qs_heatflux}
    \overline{Q} = \frac{\sum_{n=1}^{Nt} \tilde{Q}}{Nt}
\end{equation}

for different outflow forcing shows a variation of $3\%$ from the reference solution. However, there is certainly scope to improve upon this forcing and design a more robust method using traction boundary conditions.} 
\begin{figure}[h]
    \centering
    \includegraphics[width=0.8\textwidth]{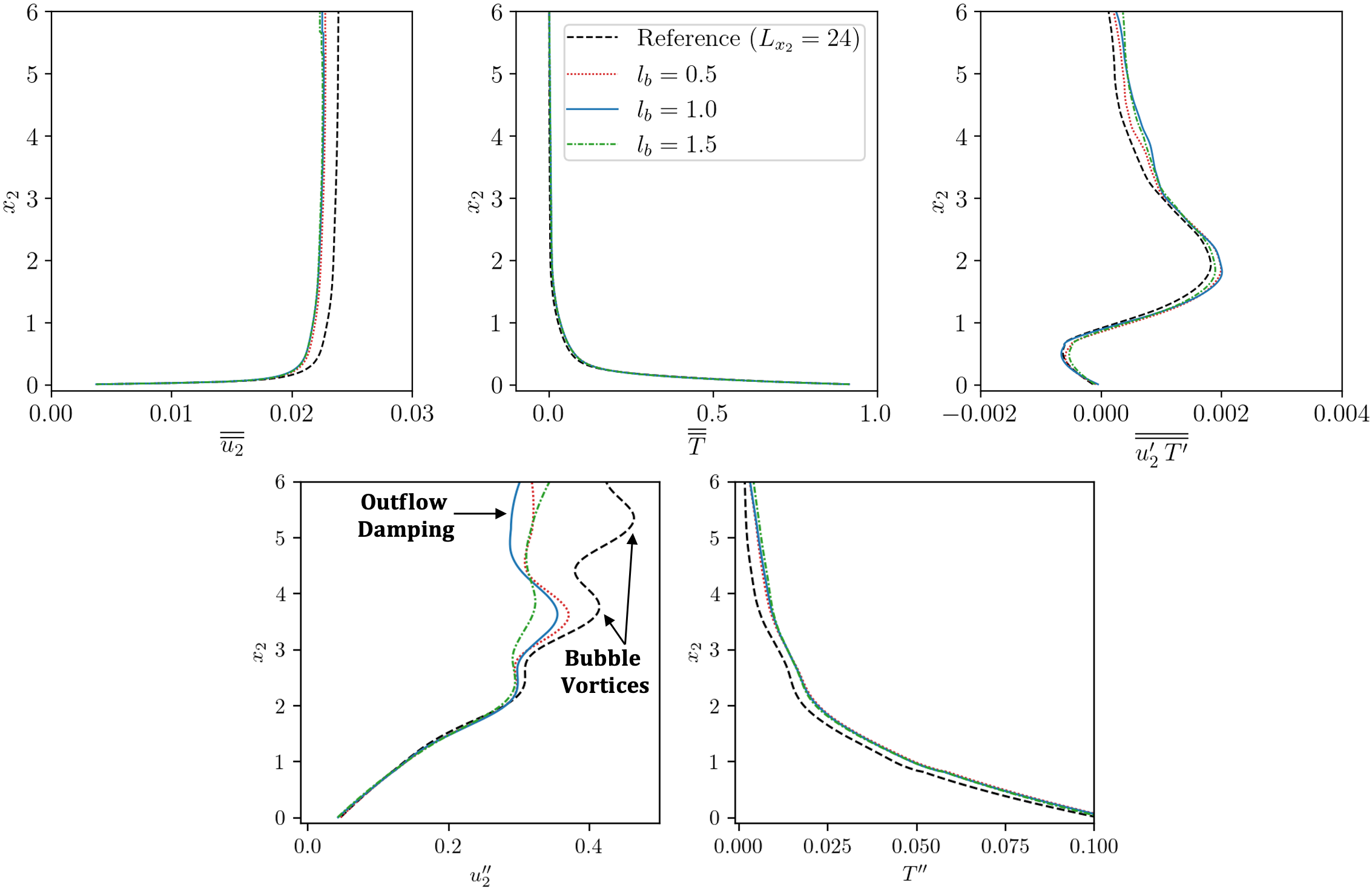}
    \caption{{ Spatio-temporal statistics for the single bubble pool boiling problem showing differences in mean and fluctuations for velocity normal to heater surface ($u_2$) and temperature ($T$), along with turbulent heat flux ($\overline{\overline{u_2'T'}}$).}}
    \label{fig:pool_boiling_statistics}
\end{figure}

{The effect of convection term in Equation \ref{eq:vel-outflow} is described in Figure \ref{fig:pool_boiling_advection} for the initial cycle of bubble growth and departure, which demonstrates that the convection term improves the quality of the outflow forcing in comparison to the solution where the contribution of this term is zero.}

{ Figure \ref{fig:pool_boiling_statistics} provides a description of spatio-temporal statistics for this problem which gives an overview of the upstream effects of the proposed outflow forcing. The statistics represent the mean and fluctuations which are described using following equation for a physical variable $f(x_1,x_2,t)$,
\begin{equation}
  \overline{f} = \frac{\sum_{n=1}^{Nt} f}{Nt} \quad\quad\quad \overline{\overline{f}} = \frac{\sum_{n=1}^{N_{x_1}}\sum_{n=1}^{Nt} f}{Nt\:{N_{x_1}}} 
\end{equation}

Here $\overline{f}$ is mean over time and $\overline{\overline{f}}$ represents mean over time and spatial direction $x_1$. Trends for velocity ($\overline{\overline{u_2}}$) and temperature ($\overline{\overline{T}}$) show very negligible differences when compared to the reference solution from the longer domain. Additionally, the turbulent heat flux ($\overline{\overline{u_2'T'}}$), which is an important metric also collapses on the same trend line. The significance of turbulent heat flux is discussed at length in our previous work \cite{DHRUV2021}. The velocity and temperature fluctuations that make up the turbulent heat flux are given by,
\begin{equation}
  f' = f - \overline{f} \quad\quad\quad f'' = \sqrt{\frac{\sum_{n=1}^{N_{x_1}}\sum_{n=1}^{Nt} (f-\overline{f})^2}{Nt\:{N_{x_1}}}}
\end{equation}

The trends for $u_2''$ and $T''$ which represent the fluctuations in velocity and temperature over space and time show the most discrepancy. The velocity fluctuations ($u_2''$) quantify the effect of the outflow forcing on the rising bubble vortices which causes significant deviation from the reference solution. This trend is expected due to the damping nature of the forcing. The upstream solution near the heater wall remains unaffected and the overall trend for turbulent heat flux make this an effective strategy for treating outflows.
}

\subsection{Two dimensional pool boiling simulations with multiple nucleation sites}

For this study, we performed a gravity scaling analysis and compared it against the heat flux model proposed by Raj { et al.} \cite{Raj2012,Raj}. Multiple nucleation sites are calculated using a Halton sequence based on the approach in our previous work \cite{DHRUV2019,DHRUV2021}. A total of $30$ nucleation sites are seeded using the nucleation site model discussed above. {The site distribution is obtained from our previous work and is scaled from 3D to 2D for the present study}. The study parameters are provided in Table \ref{table:multiple-bubble} and the resolution, $\Delta x_i = 0.03$.
\begin{figure}[h]
    \centering
    \includegraphics[width=0.95\textwidth]{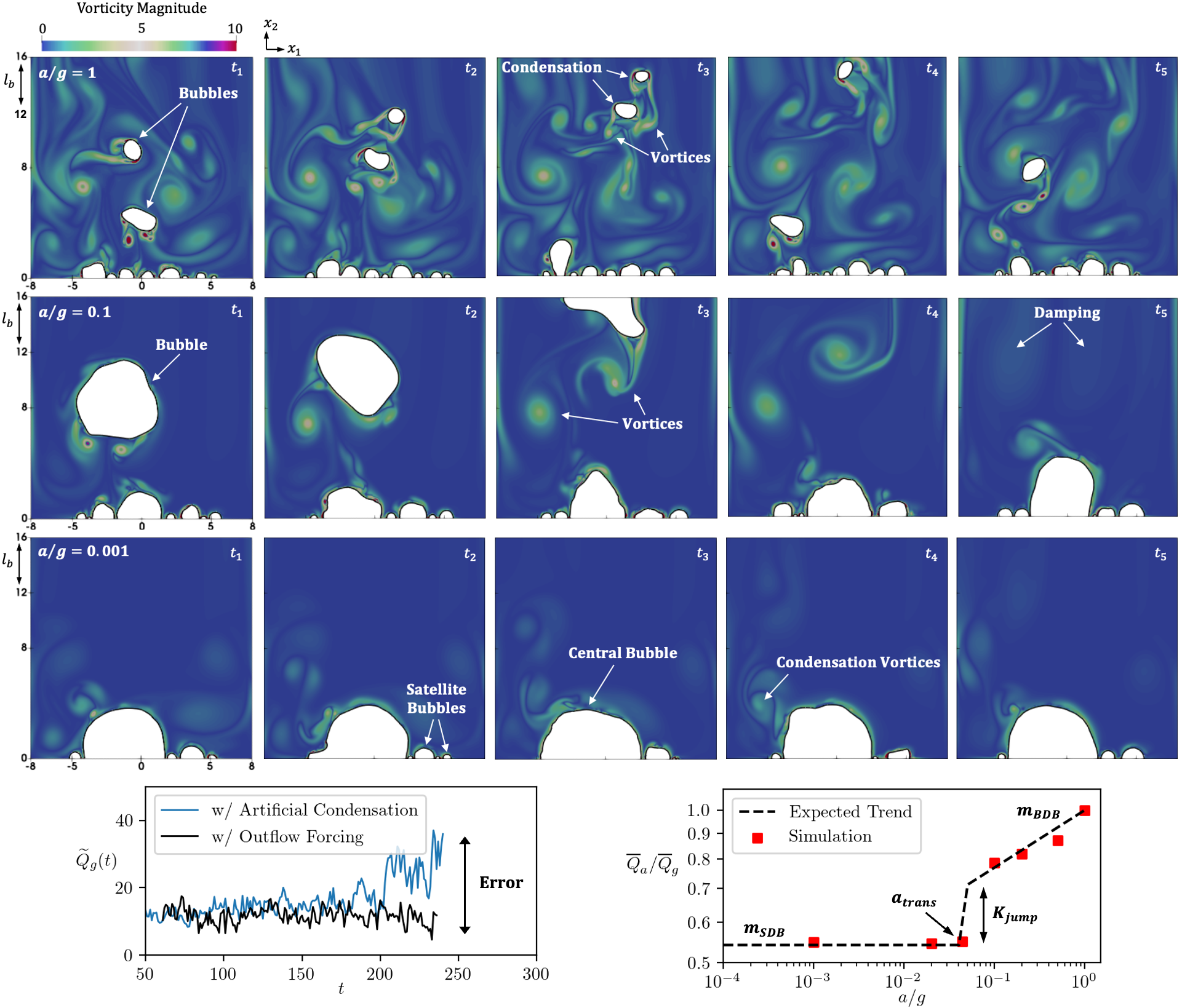}
    \caption{Vorticity contours during quasi-steady for different gravity levels, $t_1 < t_2 < t_3 < t_4 < t_5$ (top), comparison of variation in $\tilde{Q}$ on earth gravity for two different simulations using only artificial condensation versus outflow forcing described in this paper (bottom-left), and the gravity scaling of quasi-steady state heat flux, $\overline{Q}$ (bottom-right).}
    \label{fig:pool_boiling_scaling}
\end{figure}
\begin{table}
\begin{center}
\begin{tabular}{|c|c|c|c|c|c|c|c|c|c|c|c|c|c|}
\hline
Parameter & $ \rho_G/\rho_L$ & $\mu_G/\mu_L$ & $C_{p_G}/C_{p_L}$ & $k_G/k_L$ & $\alpha_G/\alpha_L$ & $\text{Re}$ & $\text{Pr}$ & $\text{St}$ & $\text{We}$ & $\psi_r$ & $\psi_a$ & $t_{wait}$ & $T_{sat}$ \\
\hline
Values       & $0.0083$ & $1$ & $0.83$ & $0.25$ & $36$ & $238$ & $8.4$ & $0.5298$ & $1$ & $45\degree$ & $90\degree$ & 0.4 & 0.15\\ \hline
\end{tabular}
\caption{Parameters for pool boiling with multiple nucleation sites \label{table:multiple-bubble}}
\end{center}
\end{table}

\begin{figure}[h]
    \centering
    \includegraphics[width=0.5\textwidth]{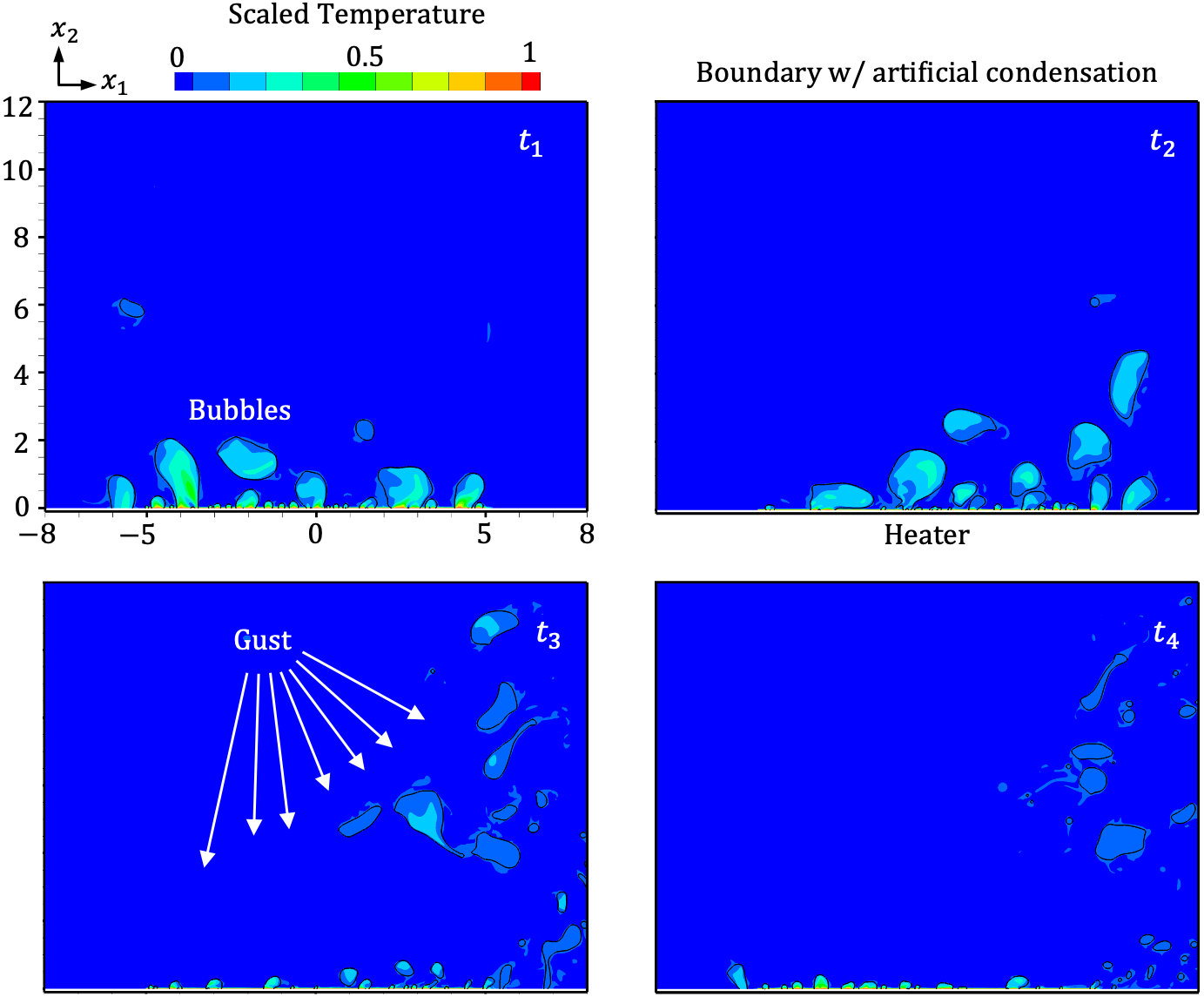}
    \caption{Schematic of non-physical gusts introduced by the artificial condensation method when applied to production problem of pool boiling involving multiple bubbles ($t_1<t_2<t_3<t_4$). This numerical artifact leads to errors in measurement of wall heat flux.}
    \label{fig:ac_schematic}
\end{figure}

Figure \ref{fig:pool_boiling_scaling} provides details of the computational domain described in the schematic in Figure \ref{fig:schematic} with the heater located between $x_1=-5$ and $x_1=5$. {To demonstrate the effectiveness of the proposed formulation we show the evolution of $\tilde{Q}$ for two different simulations using only artificial condensation versus the outflow forcing described in Section \ref{sc:outflow-forcing}. Simulation using artificial condensation near the outflow deviates over time from the quasi-steady behavior, resulting in an error in the heat flux calculation. This error is caused by non-physical gusts that appear in the numerical solution due to limitations of the artificial condensation formulation. Figure \ref{fig:ac_schematic} provides a schematic of this gust event from time time $t_1$ to $t_4$},

To compare results with previously reported trends, we performed a range of simulations for different gravity values, $Fr = 1 - 100$. The scaling, $a/g = \frac{1}{Fr^2}$, scales gravity values with respect to the value on Earth gravity, $g$. Gravity separates pool boiling in two separate regimes dominated by buoyancy or surface tension. The regimes are termed buoyancy-dominated boiling (BDB) and surface tension-dominated boiling (SDB), which are separated by the transitional acceleration, $a_{trans}$, given by
\begin{equation} \label{eq:atrans}
    a_{trans} = \frac{4.41}{Bo \: L^2_h}
\end{equation}
where, $L_h=10$, is the length of the heater and $Bo=1$, is the Bond number for the simulation. Equation \ref{eq:atrans} was identified by Raj { et al.} \cite{Raj,Raj2012} from the experimental data they collected in varying gravity environments. { The choice of $l_b$ is governed by capillary length associated with $a_{trans}$ which determines the size of disturbances that will exit of the computational domain.}

In BDB, the decrease in gravitational acceleration increases the size of departing bubbles and reduces the strength of turbulent vortical structures associated with movement and condensation of smaller bubbles within the domain, see comparison between vorticity magnitudes for $a/g=1$ and $a/g=0.1$ in Figure \ref{fig:pool_boiling_scaling}. The scaling of quasi-steady heat flux, $\overline{Q}$ (Equation \ref{eq:qs_heatflux}) with respect to gravity, $\overline{Q}_a/\overline{Q}_g$, follows the slope $m_{BDB}$.

In SDB, the dynamics are dominated by the presence of a central bubble that does not depart the heater surface, but rather acts as a vapor sink for smaller satellite bubbles, as shown in Figure \ref{fig:pool_boiling_scaling} for $a/g=0.001$. Heat flux drops sharply by a value $K_{jump}$, which is dependent on the size of the central bubble \cite{DHRUV2021}, and the slope, $m_{SDB}=0$.

The gravity scaling of heat flux computed from the simulations accurately matches the expected trend of the model proposed by Raj { et al.} \cite{Raj,Raj2012}, and serves as validation of the outflow forcing. 

\subsection{Performance evaluation for three-dimensional flow boiling simulations} \label{sc:flow-boiling-results}
{The final case we present is a demonstration of the capability of our formulation to deal with flow boiling problems that involve transport of bulk liquid through the domain, and provide detailed performance analysis of various components of the formulation. The schematic is shown in Figure \ref{fig:schematic} a, and the corresponding input values are shown in Table \ref{table:flow-boiling}, with resolution, $\Delta x_i = 0.03$.

\begin{table}
\begin{center}
\begin{tabular}{|c|c|c|c|c|c|c|c|c|c|c|c|c|c|c|}
\hline
Parameter & $ \rho_G/\rho_L$ & $\mu_G/\mu_L$ & $C_{p_G}/C_{p_L}$ & $k_G/k_L$ & $\alpha_G/\alpha_L$ & $\text{Re}$ & $\text{Pr}$ & $\text{St}$ & $\text{Fr}$ & $\text{We}$ & $\psi_r$ & $\psi_a$ & $t_{wait}$ & $T_{sat}$ \\
\hline
Values       & $0.0083$ & $1$ & $0.83$ & $0.25$ & $36$ & $238$ & $8.4$ & $0.5298$ &  $1$ & $1$ & $45\degree$ & $90\degree$ & 0.4 & 0.15 \\ \hline
\end{tabular}
\caption{Parameters for flow boiling \label{table:flow-boiling}}
\end{center}
\end{table}
The simulations are run with the bulk liquid velocity $U_b=u_0$. Figure \ref{fig:flow_boiling_performance} shows the dynamics of the flow in earth gravity that is highly bubbly and turbulent. We performed a weak scaling study for this problem by increasing the size of the computational domain in proportion to the number of processes (ranks), to maintain consistent loading of 10 AMR blocks per rank. The experiments were performed on Summit\cite{summit}, which is a leadership computing facility located at Oak Ridge National Laboratory. The reproducibility capsule and data for this problem is hosted at \cite{flow_boiling_performance}.
\begin{figure}[h]
    \centering
    \includegraphics[width=0.9\textwidth]{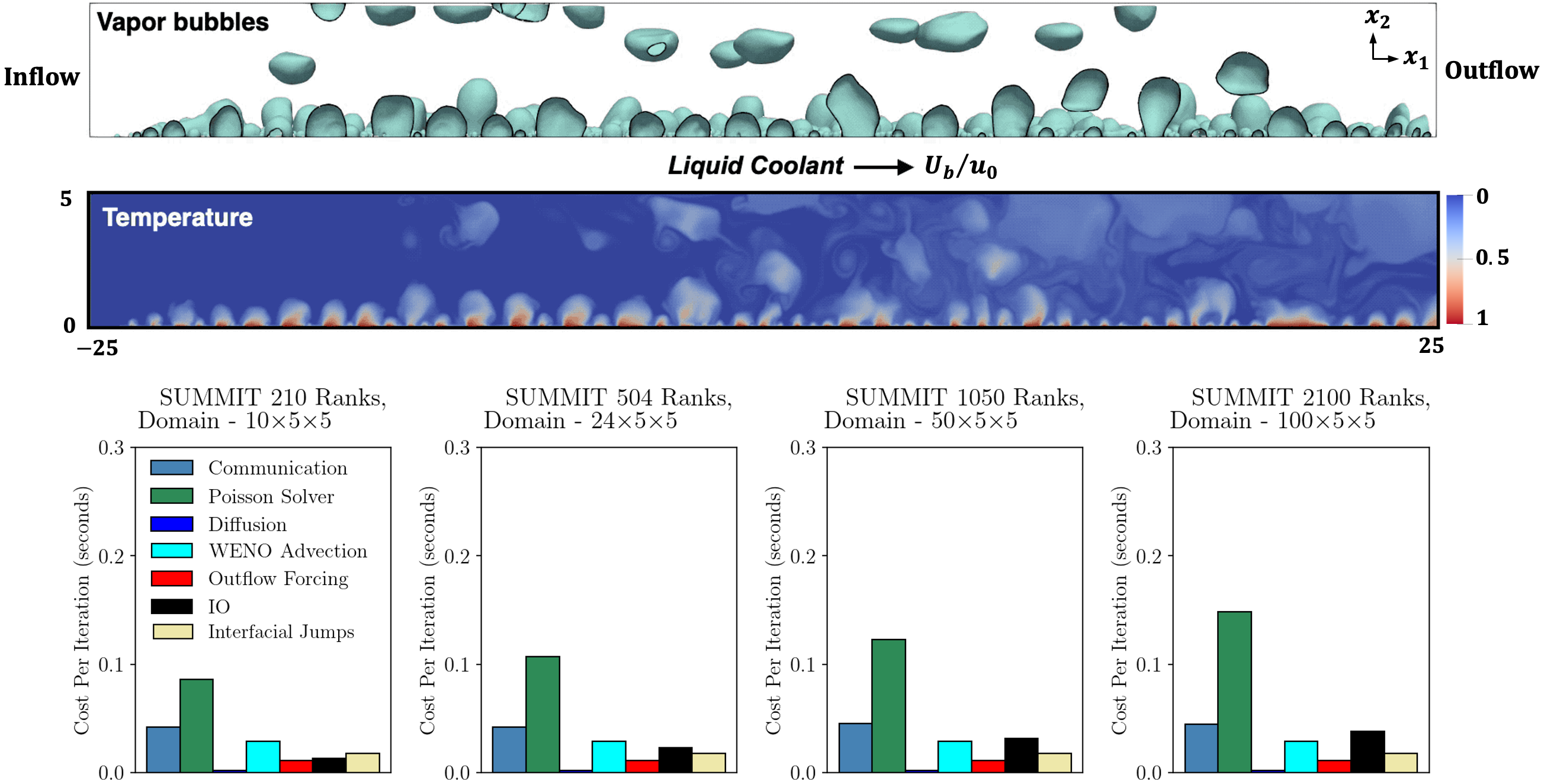}
    \caption{Three dimensional simulation of flow boiling on earth gravity, $a/g=1$, showing bubble dynamics (top), temperature distribution (middle), and weak scaling performance of different components of the solver (bottom) \cite{flow_boiling_performance}.}
    \label{fig:flow_boiling_performance}
\end{figure}

Each simulation was run for approximately $7000 s$ of wall time and the performance was logged using native implementation of timers in \flashx. Figure \ref{fig:flow_boiling_performance} shows cost per iteration for each run. As is the case with many incompressible Navier-Stokes simulations, the solution of Pressure Poisson Equation (PPE) is the highest contributor to the cost, followed by communication that is triggered by halo exchanges during guard-cell filling events. Computation of WENO advection terms for Equation \ref{eq:transport} and \ref{eq:levelset-convection} comes in third as a highest contributor on lower ranks. WENO advection stencils typically involve conditional statements that interfere with loop unrolling and optimization which contributes towards its poor computational performance. On higher ranks, Input/Output (IO) events dominate over computation of WENO advection. This is attributed to the increase in the size of domain which in-turn increases the size of data that is written to the parallel file system.

Outflow forcing formulation proposed in this work shows good performance in comparison to other components of the algorithm, and costs slighlty less than computation of interfacial jumps. Overall, the multiphase solver in \flashx can compute approximately $10 ms$ of physical time in $3600 s$ of wall time for three-dimensional flow boiling problems. This is a significant improvement over its predecessor \flash \cite{DHRUV2019}. Table \ref{table:flow-boiling-times} provides the metrics for each run.
\begin{table}
\begin{center}
\begin{tabular}{|c|c|c|c|}
\hline
Ranks & SimTime (Non-Dimensional) & SimTime ($ms$) & Wall Time ($s$) \\
\hline
210 & 1.6 & 16 & 3600\\
\hline
504 & 1.2 & 12 & 3600\\
\hline
1050 & 1.1 & 11 & 3600\\
\hline
2100 & 1.0 & 10 & 3600\\
\hline
\end{tabular}
\caption{Weak scaling of physical time per wall time for three-dimensional flow boiling simulation \label{table:flow-boiling-times}}
\end{center}
\end{table}
}
%
%
\section {Conclusion} \label{sc:conclusion}
We present an outflow forcing treatment for multiphase flow simulations involving phase transition and sharp interfacial jumps. The forcing term is designed for a fractional step predictor-corrector formulation of incompressible Navier-Stokes equations, that is coupled with a fixed pressure outflow boundary condition to enable stable exit of multiphase disturbances from the domain boundary. The forcing is applied within a buffer region near the outflow to velocity during the predictor step and to pressure jumps during the solution of pressure Poisson equation. 

The concept of an effective outflow velocity is introduced which is used as a parameter for damping and convection of bubble induced vortices, and is calculated based on the characteristic flow velocity and mean velocity for each phase within the outflow region. This strategy can be easily extended to design outflow treatment for different numerical implementations of multiphase flow solution. 

We demonstrate the applicability and accuracy of this formulation by providing results from various homogeneous and heterogeneous evaporation problems with low Weber numbers that result in large instabilities at the outflow if not handled properly. We conduct a systematic study on sensitivity of the numerical solution based on length of the buffer region, and show that the outflow forcing maintains numerical stability whilst preserving statistical behavior of the solution. We also perform a gravity scaling study for subcooled pool boiling and compare the results with heat flux models developed from experimental data to show the ability of our solver to predict different bubble regimes and their transition accurately. Finally, we show how the new outflow forcing performs during simulation of high-fidelity flow boiling problems.

{Since the proposed solution introduces damping terms in the solution of multiphase system they do not perform very well on severely truncated domains. This presents an opportunity to design more robust traction boundary conditions for these class of simulations. We hope to build on existing methods and address these challenges in our future work.}

\bigskip\noindent
\textbf{Acknowledgements:} 
The author would like to thank Dr. Anshu Dubey (Argonne National Laboratory), Dr. Klaus Weide (University of Chicago), and the reviewers for providing constructive feedback while developing the manuscript.

The material is based upon work supported by Laboratory Directed Research and Development (LDRD) funding from Argonne National Laboratory, provided by the Director, Office of Science, of the U.S. Department of Energy under contract DE-AC02-06CH11357 and the Exascale Computing Project (17-SC-20-SC), a collaborative effort of the US Department of Energy Office of Science and the National Nuclear Security Administration.

We also acknowledge that The City of Chicago is located on land that is and has
long been a center for Native peoples. The area is the traditional homelands of
the Anishinaabe, or the Council of the Three Fires: the Ojibwe, Odawa, and Potawatomi Nations.
Many other Nations consider this area their traditional homeland, including the Myaamia, Ho-Chunk, Menominee, Sac and Fox, Peoria, Kaskaskia, Wea, Kickapoo, and Mascouten.

The submitted manuscript has been created by UChicago Argonne, LLC,
operator of Argonne National Laboratory (“Argonne”). Argonne, a
U.S. Department of Energy Office of Science laboratory, is operated
under Contract No. DE-AC02-06CH11357. The U.S. Government retains for
itself, and others acting on its behalf, a paid-up nonexclusive,
irrevocable worldwide license in said article to reproduce, prepare
derivative works, distribute copies to the public, and perform
publicly and display publicly, by or on behalf of the Government.  The
Department of Energy will provide public access to these results of
federally sponsored research in accordance with the DOE Public Access
Plan. http://energy.gov/downloads/doe-public-access-plan.

\bibliographystyle{acm}
\bibliography{References}

\end{document}